\documentclass[twocolumn,amsmath,amssymb,pra]{revtex4}

\usepackage{graphics}
\usepackage{epsfig}
\usepackage{amsmath}
\usepackage{color}
\usepackage{dcolumn}
\usepackage{bm}

\newcommand{\ds }{\displaystyle}

\begin{document}

  \bibliographystyle{apsrev}

  \title{Influence of a tight isotropic harmonic trap on 
    photoassociation in ultracold homonuclear alkali gases}

       \author{Sergey Grishkevich and Alejandro Saenz}

       \affiliation{AG Moderne Optik, Institut f\"ur Physik,
         Humboldt-Universit\"at zu Berlin, Hausvogteiplatz 5-7,
         10117 Berlin, Germany}

       \date{\today}

  \begin{abstract}

     The influence of a tight isotropic harmonic trap on photoassociation 
     of two ultracold alkali atoms forming a homonuclear diatomic is 
     investigated using realistic 
     atomic interaction potentials. Confinement of the initial atom pair due 
     to the trap leads to a uniform strong enhancement of the photoassociation 
     rate to most, but also to a strongly suppressed rate for 
     some final states. Thus tighter traps do not necessarily enhance the 
     photoassociation rate. A further massive enhancement of the
     rate is found for strong interatomic interaction potentials. The 
     details of this interaction play a minor role, except for large 
     repulsive interactions for which a sharp window occurs in the 
     photoassociation spectrum as is known from the trap-free case.  
     A comparison with simplified models 
     describing the atomic interaction like the pseudopotential approximation 
     shows that they often provide reasonable estimates for the trap-induced 
     enhancement of the photoassociation rate even if the predicted rates 
     can be completely erroneous. 

  \end{abstract}

  \maketitle

%---------------------------
\section{Introduction}
%---------------------------
%
\label{sec:intro}

  Over the past ten years there has been an increasing interest in ultracold 
  atomic and molecular physics. This interest was stimulated by the 
  successful experimental observation of Bose-Einstein condensation (BEC) in 
  dilute atomic gases~\cite{cold:ande95}. These atomic condensates exhibit 
  many qualitatively new features. Besides their relevance to fundamental 
  quantum statistical and possibly even solid-state questions a further 
  interesting aspect is that the atoms can bind together to form ultracold and
  even Bose-Einstein condensed 
  molecules~\cite{cold:joch03,cold:rega03,cold:zwie03}. 

  Although so far the only successful way for achieving a molecular 
  BEC is based on magnetic Feshbach resonances, alternative schemes are still
  highly desirable, since magnetic Feshbach resonances do not appear to be a 
  universal tool. One of the alternative schemes is photoassociation where two 
  ultracold or Bose-condensed atoms absorb a photon and form a bound excited 
  molecule~\cite{cold:lett93,cold:fior98}. Although it was demonstrated that 
  this process generates cold molecules, the yield is small compared
  to the one obtained by means of magnetic Feshbach resonances. The advantage
  of photoassociation (and related coherent-control schemes) compared to 
  Feshbach resonances is, however, their assumed wider range of applicability,
  since there is no need for the occurrence of suitable resonances and 
  thus no requirement for specific magnetic properties of the atoms 
  involved.

  Besides simple one-step photoassociation that yields electronically excited
  molecules there are also resonant or non-resonant multi-step schemes 
  leading to the electronic and possibly even rovibrational ground state. 
  One of the schemes to produce molecules from atoms with help of lasers 
  is, e.\,g., two-color stimulated Raman adiabatic
  passage~\cite{cold:drum02}. 
  The present work discusses only one-photon association 
  explicitly, but it is important to note that the discussed  
  transition matrix elements are direct ingredients for the modeling 
  of more sophisticated schemes like the mentioned two-photon 
  processes. 

  Photoassociation is
  also a powerful tool for the investigation of the properties of cold atoms
  and diatomic molecules. The absorption of the photon typically occurs at 
  large internuclear distances, and thus the photoassociation spectrum 
  provides important information about the long-range part of the molecular 
  potential curves as well as the collisional properties of 
  atoms~\cite{cold:abra95, cold:tiem96, cold:fior01, cold:gutt02}.

  The cooling of atomic samples is usually achieved in a trap and thus
  photoassociation experiments in ultra-cold atomic gases are performed in
  the presence of a trap potential. In most cases these traps are rather 
  shallow so that the corresponding trap frequency $\ds \omega$ is of the 
  order of 100\,Hz~\cite{cold:schl03}. (Here and in the following the
  frequency of the trap corresponds to the one in the harmonic 
  approximation). For such a frequency the influence of the trap is expected 
  to be negligible. This may, however, change for very tight traps. In fact, 
  it was pointed out that the atom-molecule conversion process is more
  efficient, if photoassociation is performed under tight trapping
  conditions as they are, e.\,g., accessible in optical 
  lattices~\cite{cold:jaks02}.
  The advantage of using tight confinement has stimulated further theoretical
  investigations and very recently some proposals were made that discuss the 
  possibility of using the trapping potential itself for the formation of 
  molecules~\cite{cold:bold05,cold:koch05}.

  The study of photoassociation in tight optical lattices is of interest by 
  itself, since it is possible to achieve tailored Mott-Insulator states 
  containing a large number of almost identical lattice sites, each filled 
  with exactly two atoms~\cite{cold:grei02}. The trap frequency of a 
  lattice site in which molecules are produced via photoassociation can be 
  of the order of 100\,kHz~\cite{cold:rom04}. The systematic investigation 
  of the influence of a tight isotropic harmonic trap on the photoassociation 
  process of two alkali atoms forming a homonuclear molecule is the 
  topic of this work. Realistic atom-atom interaction potentials are adopted.
  This allows to check also the range of applicability of the 
  $\delta$-function (pseudopotential) approximation for the description of 
  the photoassociation process.

  In the pseudopotential approximations the true atom-atom interaction is 
  replaced by one that reproduces asymptotically the two-body zero-energy 
  s-wave scattering. 
  For this choice of the potential ($\delta$-function) and if the two atoms 
  are placed in a harmonic trap the Schr\"odinger equation 
  possesses an analytical solution~\cite{cold:busc98,cold:idzi05}. The
  validity regime of the pseudopotential approximation has been discussed
  with respect to the energy levels for trapped atoms in
  \cite{cold:blum02}. It was shown that the use of an
  energy-dependent pseudopotential (instead of the mostly adopted
  energy-independent one) gives almost correct energy levels for
  two harmonically trapped atoms. Whether this simplified model for the
  atomic interaction is appropriate for the description of photoassociation in
  a harmonic trap is, however, not immediately evident. Therefore, the present
  work compares the results obtained using realistic atomic interaction
  potentials with the ones obtained with either the energy-dependent or
  -independent pseudopotential.

  Photoassociation in tight traps has been studied theoretically 
  before \cite{cold:deb03}. The energy-independent 
  pseudopotential approximation was adopted and only photoassociation into 
  long-range states discussed. Since the present work uses realistic 
  atomic interaction potentials, transitions to all final vibrational states 
  can be considered. This allows to identify two different regimes with respect
  to the influence of a tight trap on the photoassociation rate as well as 
  (approximate) rules where a transition from one regime to the other is to 
  be expected. 

  The outline of this work is the following. First, a brief description of 
  the model systems is given in Sec.\,\ref{sec:system}. In 
  Sec.\,\ref{sec:photoassociation} the influence of a tight trap on 
  photoassociation is discussed. This includes after a brief general 
  discussion of the trap influence in Sec.\,\ref{sec:spectrum} 
  the derivation of a sum rule in Sec.\,\ref{sec:sumrule}, 
  the introduction of an enhancement or suppression factor in 
  Sec.\,\ref{sec:ratio}, and the discussion of two regimes 
  in Sec.\,\ref{sec:constregim} and \ref{sec:cutoffregim}. 
  Then the case of repulsive atom-atom interactions is considered in 
  Sec.\,\ref{sec:positivesclen}. The combined influence of trap 
  frequency and atom-atom interaction on photoassociation is 
  investigated in Sec.\,\ref{sec:combined}, the validity of 
  the pseudopotential approximation in Sec.\,\ref{sec:delta}. 
  Finally, a discussion and outlook is given in 
  Sec.\,\ref{sec:discussion}. 
  All equations and quantities in this paper are given in atomic units unless 
  otherwise specified.

%-------------------
\section{The system}
%-------------------
%
\label{sec:system}

   Photoassociation of two identical atoms confined in an isotropic harmonic 
   trap and interacting through a two-body Born-Oppenheimer potential 
   $\ds V_{\rm int}(R)$ is considered. The spherical symmetry and 
   harmonicity of 
   the trap allows to separate the center-of-mass and the radial internal 
   motion~\cite{cold:busc98}. The eigenfunctions of the center-of-mass motion 
   are the harmonic-oscillator states. Thus the problem reduces to solving the 
   Schr\"odinger equation for the radial internal motion %
     \begin{eqnarray}
       \ds
       \left[ \frac{1}{2\mu}\frac{d^2}{dR^2}\,
             -\,\frac{J(J+1)}{2\mu R^2} \,\right.&-& V_{\rm int}(R)\,
             -\, \frac12\mu\omega^2R^2  \nonumber \\
                                   &+& \left. E \, \right]\,\Psi (R)\,=\,0\; .
       \label{SE_rim}
     \end{eqnarray}
   In Eq.~(\ref{SE_rim}) $J$ denotes the rotational quantum number, 
   $\ds \omega$ is the harmonic trap frequency, and $\ds \mu $ is the reduced 
   mass that is equal to $m/2$ in the present case of particles with identical 
   mass $m$.

   In order to compute the photoassociation spectrum the vibrational wave 
   functions $\Psi(R)/R$ are determined for the initial and final molecular 
   states from Eq.~(\ref{SE_rim}) with the corresponding Born-Oppenheimer 
   interaction potentials $V_{\rm int}(R)$. The equation is solved 
   numerically using an expansion in $B$ splines. 
   For the investigation of the influence of the trap on the 
   photoassociation rate Eq.~(\ref{SE_rim}) is solved for $\omega\neq 0$.  

   The photoassociation processes most relevant to experiments on ultracold 
   alkali atoms correspond to transitions from two free 
   ground-state atoms (interacting {\it via} the ground triplet or singlet  
   potential) to the different vibrational levels of the 
   first excited triplet or singlet 
   state~\cite{cold:alma99,cold:alma01,cold:kemm04}. Due to hyperfine 
   interaction, two alkali atoms can also interact via a coherent admixture 
   of singlet and triplet states. This work starts by considering    
   the photoassociative transition between the two triplet states 
   $a ^3\Sigma^+_u$ and $1 ^3\Sigma^+_g$ for $^6 \mbox{Li}$. 
   A corresponding experiment is, e.\,g., reported in~\cite{cold:schl03}. 
   The generality of the conclusions drawn from this specific example are 
   then tested by considering also other atoms ($^7$Li and $^{39}$K) or 
   modifying artificially the interaction strength, as is discussed below. 

   For the short-range part of the $a ^3\Sigma^+_u$ molecular potential of 
   Li$_2$ the data in~\cite{cold:cola03} are used, including the van der Waals 
   coefficients cited therein. In the case of the $1 ^3\Sigma^+_g$ state 
   data for interatomic distances between $\ds R=4.66\,a_0$ and 
   $\ds R=7.84\,a_0$ 
   are taken from~\cite{cold:lint89} and are extended with {\it ab initio} 
   values from~\cite{cold:schm85} for distances between $\ds R=3.25\,a_0$ 
   and $\ds R=4.50\,a_0$ and between $\ds R=8.0\,a_0$ and $\ds R=30.0\,a_0$. 
   The van der Waals coefficients from~\cite{cold:mari95} 
   are used. For a $\Sigma$ to $\Sigma$ molecular dipole transition 
   the selection rule is $J=J'\pm 1$. Assuming ultracold atomic 
   gases the atoms interact initially in the $\ds J'=0$ state of 
   the $a ^3\Sigma^+_u$ potential. The dipole selection rule leads then to 
   transitions to the $\ds J=1$ states of $1 ^3\Sigma^+_g$. With the given 
   potential-curve parameters a solution of Eq.~(\ref{SE_rim}) 
   in the absence of a trap ($\omega=0$) yields for the fermionic 
   $^6$Li atoms 10 and 100 vibrational 
   bound states for the $a ^3\Sigma^+_u$ ($J'=0$) and the $1 ^3\Sigma^+_g$ 
   ($J=1$) states, respectively. In the case of the bosonic $^7$Li atoms 
   there are 11 and 108 vibrational 
   bound states for the $a ^3\Sigma^+_u$ ($J'=0$) and the $1 ^3\Sigma^+_g$ 
   ($J=1$) states, respectively.

   The electronic dipole moment $D(R)$ for the transition  
   $a\, ^3\Sigma^+_u \rightarrow 1\, ^3\Sigma^+_g$ of Li was 
   calculated~\cite{cold:vanne} with a configuration interaction (CI) method 
   for the two valence electrons using the code described 
   in~\cite{bsp:vann04}. The core electrons were described with the aid of 
   the Klapish model potential with the parameters given 
   in~\cite{aies:magn99} 
   and polarization was considered as discussed in~\cite{bsp:dumi06}. 
   The resulting $D(R)$ (and its value in the separated atom 
   limit) is in good agreement with literature  
   data~\cite{cold:schm85,cold:ratc87,cold:pipi91,cold:mari95}. 

   In the limit of zero collision energy the interaction between two atoms 
   can be characterised by their s-wave scattering length $\ds a_{\rm sc}$. 
   Its sign determines the type of interaction (repulsive or attractive) 
   and its absolute value the interaction strength. For a given potential 
   curve the s-wave ($J'=0$) scattering length can be determined using the 
   fact that at large distances the scattering wave function describing the 
   relative motion (for $\omega=0$ and very small 
   collision energies) reaches an asymptotic behavior of the form % 
    \begin{equation} 
      \ds
      \Psi_{E}(R)=\sqrt{\frac{k}{\pi E}}~\sin\left[k(R-a_{\rm sc})\right]\; .
      \label{eq:asymptot_wf}
    \end{equation}
   In the present numerical approach discretised continuum states are 
   obtained, since the wave-function calculation is performed within a 
   finite $\ds R$ range, i.\,e.\ in the interval $[0,R_{\rm max}]$.  
   Only wave functions that decay before or have a node at $R_{\rm max}$ are 
   obtained. From the analysis of the lowest lying discretised continuum state 
   the scattering length $a_{\rm sc}$ is obtained by the use of the relation 
   $\ds a_{\rm sc} = R_{\rm max} - \frac{\pi}{k}$  with 
   $\ds k=\sqrt{2\mu E}$~\cite{cold:juli96}. A variation of 
   $R_{\rm max}$ changes the continuum discretization and therefore results  
   in $R_{\rm max}$-dependent lowest lying continuum solutions $\Psi_{E_0}$. 
   The scattering length extracted from $\Psi_{E_0}$ converges, however,  
   to a constant value as $R_{\rm max}$ increases and $E_0$ approaches 
   zero. Using this method and the adopted potential curves 
   the scattering length values $\ds a_{\rm sc} = -2030\, a_0$ and 
   $\ds a_{\rm sc} = -30\, a_0$ are obtained for $^6$Li and $^7$Li, 
   respectively.  
   These values agree well with the experimental ones:   
   $a_{\rm sc} = (-2160\pm 250)\,a_0$ ($^6$Li) and 
   $a_{\rm sc} = (-27.6\pm 0.5)\,a_0$ ($^7$Li)~\cite{cold:abra97}. 

   The interaction of two ultracold $^6$Li atoms is strongly, the one 
   of $^7$Li weakly attractive, as is reflected by the large and small  
   but negative scattering lengths. In the case of two identical fermionic 
   $^6$Li atoms the asymmetry requirement of the total 
   wave function excludes s-wave scattering. Thus the present results 
   are more applicable for two $^6$Li atoms in different hyperfine states 
   (where the admixture of a singlet potential would, however, usually 
    modify the scattering length), but 
   are actually meant as a realistic example for a very large negative 
   scattering length, i.\,e.\ strong attraction. In order to further check 
   the generality of the results also the formation of $^{39}$K$_2$ is 
   investigated as an example for a small repulsive interaction. 
   In this case photoassociation starting from two potassium atoms 
   interacting {\it via} the {\it singlet} $X ^1\Sigma^+_g$ ground 
   state and transitions into the $A ^1\Sigma_u^+$ state are considered. 
   This process is not only experimentally relevant \cite{cold:niko99}, 
   but is at the same time an even further check of the generality of the 
   conclusions obtained from the investigation of the transitions between 
   {\it triplet} states in $\mbox{Li}_2$.

   The data for constructing the relevant potential curves for $^{39}$K$_2$ 
   are taken from~\cite{cold:allouche,cold:amio95,cold:wang97}. 
   The resulting potential curve for the $X ^1\Sigma^+_g$ state yields a 
   scattering length $a_{\rm sc}\approx +90\,a_0$. This is in reasonable 
   agreement with the experimental value given in \cite{cold:will99} where 
   $a_{\rm sc}$ is found to be lying between $+90\,a_0$ and
   $+230\,a_0$. 

   Instead of selecting additional atomic pairs that could 
   represent examples for other values of the scattering length, the 
   sensitivity of the s-wave interaction on the position of the least 
   bound state is used to generate {\it artificially} a variable interaction 
   strength. The scattering length is thus modified 
   by a variation of the particle mass. The strong mass dependence of 
   the scattering length is already evident from its change from $-2030\,a_0$ 
   to $-30\,a_0$ for the isotopes $^6$Li and $^7$Li, respectively. 
   Experimentally, a strong variation of the interaction strength 
   can be realized by the aid of magnetic Feshbach 
   resonances~\cite{cold:loft02,cold:rega03b}.

%------------------------------------------------------------------------------
\section{Photoassociation in an isotropic harmonic trap}
%------------------------------------------------------------------------------
\label{sec:photoassociation}

%--------------------------------------------------------------------------
\subsection{Photoassociation in a trap}
\label{sec:spectrum}

\begin{figure}[ht]             
 \centering
\includegraphics[width=8.0cm,height=6.5cm]{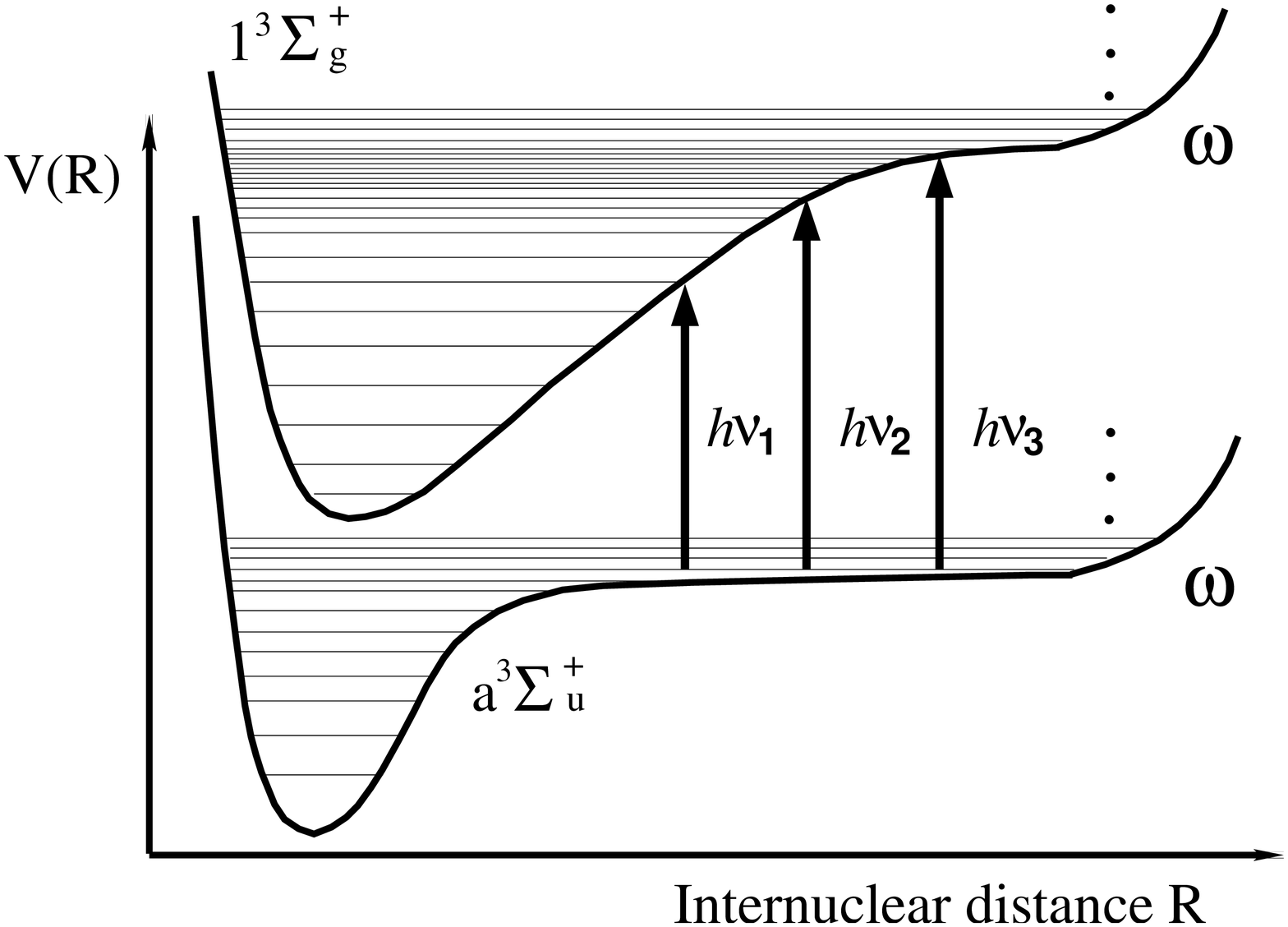}
 \caption{{\footnotesize
 Sketch of the photoassociation process for $ ^6 \mbox{Li}_2$ in the 
 presence of a trap (not to scale). With the aid of a laser 
 photoassociation is induced from the first trap-induced bound state 
 ($\ds v' = 10$) of the $a ^3\Sigma^+_u$ state into some vibrational 
 level $\ds v$ of the $1 ^3\Sigma^+_g$ state. Different laser frequencies 
 $\nu_i$ couple the same initial state to different final states.
} }\label{fig:PAscetch}
\end{figure}

 While Eq.~(\ref{SE_rim}) yields in the trap-free case ($\omega=0$) 
 both bound (vibrational) and continuum (dissociative) states, the 
 harmonic trap potential changes the energy spectrum to a purely  
 discrete one, as is sketched in Fig.\,\ref{fig:PAscetch}.
 Considering the 
 concrete example of two $^6$Li atoms where the 
 $a ^3\Sigma^+_u$ state supports the 10 vibrational bound states 
 $v'=0$ to 9, $v'=10$ ($J'=0$) denotes the first state that 
 results from the trap-induced continuum discretization. This (first 
 trap-induced) state describes the initial state of two spin-polarized 
 $^6$Li atoms interacting {\it via} the $a ^3\Sigma^+_u$ potential curve, 
 if a sufficiently cold atomic gas in an 
 (adiabatically turned-on) harmonic trap is considered. In the present 
 work photoassociation (by means of a suitably tuned laser) from this 
 initial state to one of the vibrational states $v$ of the $1 ^3\Sigma^+_g$ 
 potential is investigated as a function of the trap frequency $\omega$. 
 In view of the already discussed relevant dipole-selection rule the final 
 state possesses $J=1$ and in the following $J'=0$ and $J=1$ is tacitly 
 assumed. The strength of the photoassociation transition to final state 
 $v$ is given by the rate \cite{cold:cote98b}
\begin{equation}          
\ds
     \Gamma_{v}(\omega) = 4\pi^2 \mathcal{I}\, I^v(\omega) 
\label{PA_total}
\end{equation}
where $\mathcal{I}$ is the laser intensity and  
\begin{equation}           
\ds
 I^{v} (\omega) = \left| \int\limits_0^\infty \: \Psi^{v}(R;\omega) 
              \: D(R) \: 
              \Psi^{\rm 10'}(R;\omega) \:{\rm d}R \right|^2\; .
\label{PA_dipole}
\end{equation}
 In Eq.\,(\ref{PA_dipole}) $\Psi^v(R)/R$ and $\Psi^{\rm 10'}(R)/R$ 
 are the vibrational wave functions of the final and initial state, 
 respectively. Since the radial density is proportional to $|\Psi|^2$, 
 it is convenient to discuss $\Psi$ instead of the true vibrational 
 wave function $\Psi(R)/R$. This will be done in the following where 
 $\Psi$ is for simplicity called vibrational wave function. Finally, $D(R)$ 
 is the ($R$-dependent) electronic transition dipole matrix element between 
 the $a ^3\Sigma^+_u$ and the $1 ^3\Sigma^+_g$ state of $\mbox{Li}_2$ 
 introduced in Sec.\,\ref{sec:system}.  $D(R)$ is practically constant for 
 $R > 25\,a_0$. Eq.\,(\ref{PA_dipole}) is only valid within the dipole 
 approximation. The latter is supposed to be applicable, if the photon 
 wavelength is much larger than the extension of the atomic or molecular 
 system. The shortest photoassociation laser wavelength corresponds to 
 the transition to the highest-lying vibrational state and is thus 
 approximately the one of the atomic (2\,$^2$S $\rightarrow$ 2\,$^2$P 
 transition), $\lambda = 12680\,a_0$. Although the spatial extent of some 
 of the final vibrational states (and of course the initial state in the 
 case of shallow traps) has a similar or even larger extent, beyond 
 dipole approximation effects are neglected in this work.  
 
 The key quantity describing the photoassociation rate 
 to different vibrational states $v$ or for variable trap frequency $\omega$ 
 is $I^{v} (\omega)$ on whose calculation and discussion this 
 work concentrates. It is important to note that also in the case of more 
 elaborate laser-assisted association schemes like stimulated Raman processes 
 that involve (virtual) transitions to the $v$ states the transition rate is 
 proportional to $I^{v} (\omega)$.  

\begin{figure}[ht]     
\centering
\includegraphics[width=8.5cm,height=6.5cm]{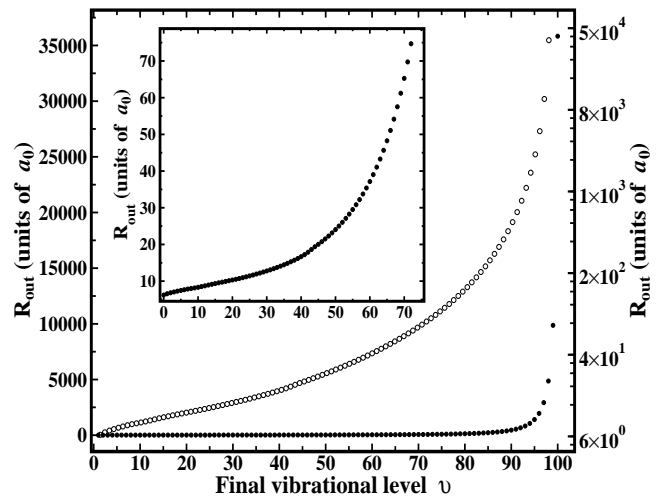}
\caption{{\footnotesize
     The classical outer turning points of the vibrational levels of the 
     $1\,^3\Sigma_g^+$ state of $^6$Li$_2$ are shown 
     on a linear (solid circles, left scale and insert) and on a 
     logarithmic scale (empty circles, right scale).
}}\label{fig:ClasTurnPoint}
\end{figure}
\begin{figure}[ht]         
 \centering
\includegraphics[width=8.5cm,height=6.5cm]{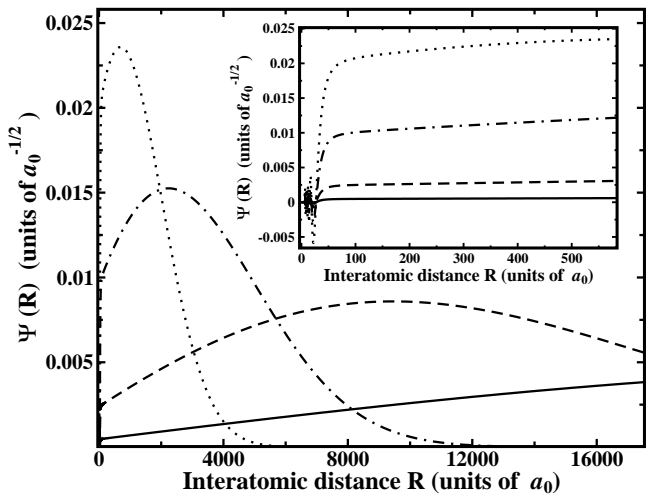}
 \caption{{\footnotesize
     Wave functions of the initial $a\,^3\Sigma_u^+$ state of 
     $ ^6 \mbox{Li}_2$ for trap frequencies
     $\omega = 2\pi\times\mbox{1kHz}$ (solid),
     $\omega = 2\pi\times\mbox{10kHz}$ (dashes),
     $\omega = 2\pi\times\mbox{100kHz}$ (chain), and 
     $\omega = 2\pi\times\mbox{500kHz}$ (dots).
     (The insert shows the small $\ds R$ range on an enlarged scale.)
} }\label{fig:iniWF4wregimes}
 \end{figure}

 According to Eq.\,(\ref{PA_dipole}) the photoassociation rate depends for 
 transitions between long-range states on the Franck-Condon factors between 
 the initial and final nuclear wave functions, if $D(R)$ is 
 practically constant for large $R$. In the case of alkali atoms the 
 interaction potentials of the electronic states can be very long 
 ranged and can contain numerous rovibrational bound states. 
 Fig.~\ref{fig:ClasTurnPoint} shows, e.\,g., the classical outer turning 
 points $R_{\rm out}$ of the 100 ($J=0$) vibrational bound states of 
 $^6$Li$_2$ supported by the final-state electronic potential curve 
 $1 ^3\Sigma^+_g$. The orthogonality of the states is achieved 
 by the occurrence of $v'$ nodes. As $v'$ increases the 
 wavefunctions consist of a highly oscillatory short range part 
 with small overall amplitude that covers the range of the $v'-1$ 
 wavefunction and a large outermost lobe. The $1 ^3\Sigma^+_g$ 
 state is very long ranged, since its leading van der Waals term  
 is $\ds -C_3/R^3$. The initial electronic state $a ^3\Sigma^+_u$ 
 with leading $\ds -C_6/R^6$ van der Waals term is shorter ranged.  
 Fig.~\ref{fig:iniWF4wregimes} shows the initial vibrational state 
 for $^6$Li as a function of the trap frequency. This first trap-induced 
 bound state possesses $v$ nodes (here $v=10$) that are located in the 
 $R$ range of the last trap-free bound state ($v=9$). The overall 
 amplitude in this about $25\,a_0$ long interval is very small and most 
 of the wavefunction is distributed over the harmonic trap.  

\begin{figure}[ht]              
 \centering
\includegraphics[width=8.5cm,height=13cm]{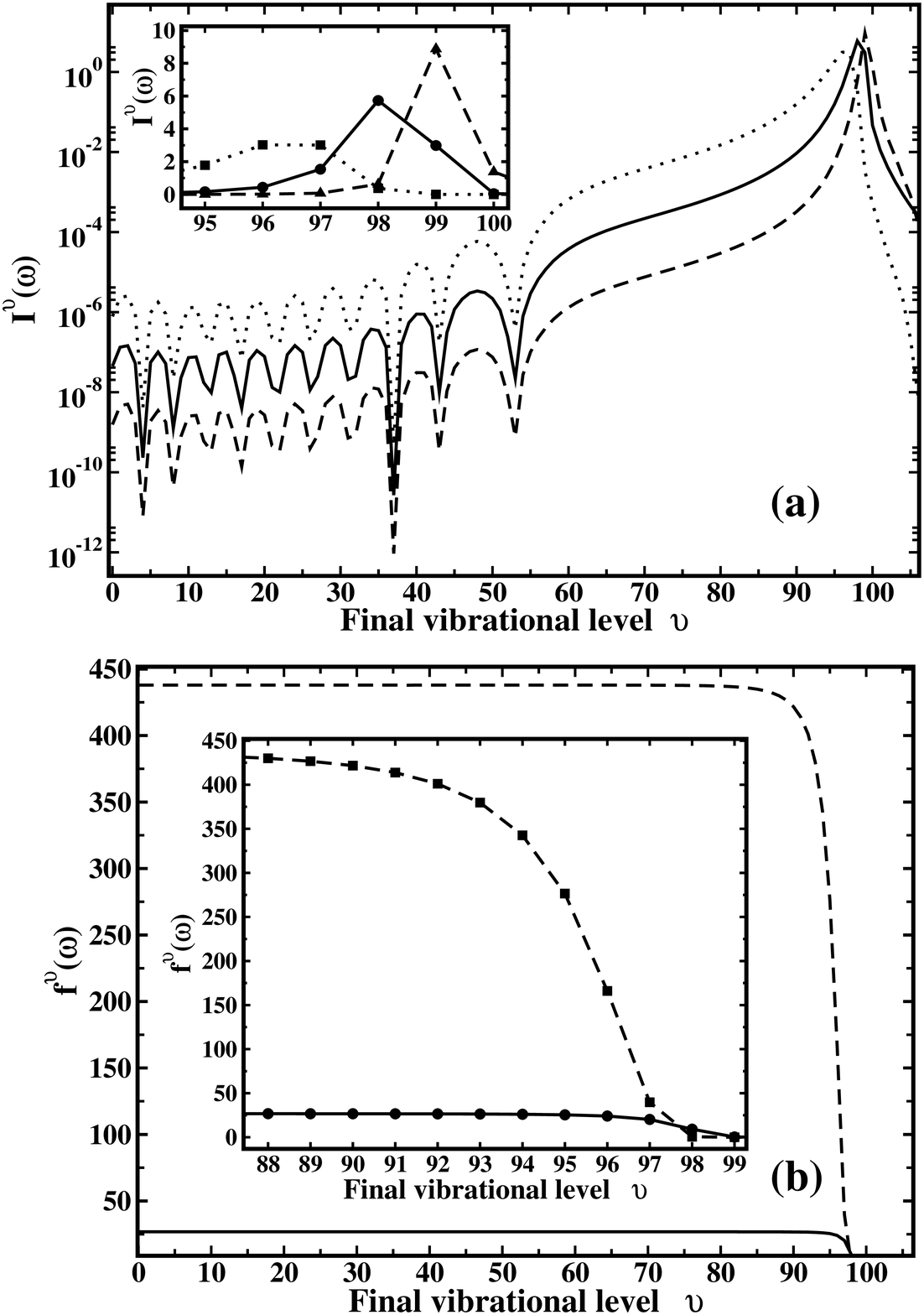}
 \caption{{\footnotesize
     (a) The squared dipole transition moments $\ds I^{v}(\omega)$  
     describing transitions from the 
     trap-induced ($v'=10$) initial $a ^3\Sigma^+_u$ 
     state to the vibrational manifold ($v$) of the
     $1 ^3\Sigma^+_g$ state of $ ^6 \mbox{Li}_2$ are shown for 
     the trap frequencies  
     $\omega = 2\pi\times 1\mbox{kHz}$ (dashes), 
     $\omega = 2\pi\times 10\mbox{kHz}$ (solid), 
     and $\omega = 2\pi\times 100\mbox{kHz}$ (dots). 
     The insert shows the transitions to $v=95$ to 100 on an 
     enlarged scale.
     For a better visibility the discrete transitions (marked 
     explicitly with different symbols in the insert) are plotted 
     as continuous lines.
     (b) The ratio $\ds f^{v}(\omega)$ [defined
     in Eq.\,(\ref{f_v})] is shown for 
     $\ds \omega = 2\pi\times 10\,\mbox{kHz}$ (solid) and 
     $\omega = 2\pi\times 100\,\mbox{kHz}$ (dashes)
     as a function of the final vibrational level $v$. 
     (As in (a) the discrete points are connected 
     by lines to guide the eye.) 
     The insert shows the transitions to $v=88$ to 99 on a 
     magnified scale.  
} }\label{fig:PAtransmomAttract}
 \end{figure}

 The squared transition dipole moments $\ds I^{v}(\omega)$
 are shown for $ ^6 \mbox{Li}$ in Fig.~\ref{fig:PAtransmomAttract}(a) 
 for three different trap frequencies $\ds \omega$. 
 As mentioned before, the final vibrational levels with $v > 99$ are 
 trap-induced bound states and exist only due to the 
 continuum discretization in the presence of a trap. If the trap 
 would be turned-off (adiabatically) after photoassociation to such 
 a level, the trap induced dimer would immediately dissociate 
 (without the need for any (radiative or non-radiative) coupling to 
 some dissociative state). 

 For a fixed trap frequency the photoassociation 
 rate generally increases as a function of the final vibrational level $v$, 
 but for small $v$ an oscillatory behavior is visible. These 
 oscillations are a consequence of the nodal structure of the initial-state 
 wave functions describing the atom pair. The 10 nodes (for the shown 
 example of $^6$Li) of the initial-state wave function lead to exactly 
 10 dips in photoassociation spectrum. Their exact position depends on 
 the interference with the nodal structure of the final-state 
 wave functions. The oscillatory structure of $I^v(\omega)$ 
 ends at about $v=55$ and beyond that point the rate 
 increases by orders of magnitude, before a sharp decrease is 
 observed close to the highest lying vibrational bound state 
 ($v=99$). The absence of oscillatory behavior is a clear signature that 
 for those transitions (in the present example for transitions into 
 states with $v>55$) the Franck-Condon factors are determined by 
 the overlap of the outermost lobe of the initial state with 
 the one of the final state.  

 The comparison of $I^v(\omega)$ for the different trap frequencies 
 shown in Fig.\,\ref{fig:PAtransmomAttract}(a) indicates a very 
 systematic trend. The transition probabilities to most of the 
 vibrational bound states increases with increasing trap frequency. 
 This is in accordance with simple confinement arguments, since a 
 tighter trap confines the atoms in the initial state to a smaller 
 spatial region. Due to the special properties of harmonic traps, this 
 confinement translates directly into a corresponding confinement of 
 the pair density (see Eq.(\ref{SE_rim})). The probability for atom 
 pairs to have the correct separation for the photoassociative 
 transition is thus expected to increase for tighter confinements, 
 since a larger Franck-Condon overlap of the now more compact initial 
 state with the bound molecular final state is expected. However, for 
 the vibrational final states close to and above the (trap-free) 
 dissociation threshold a completely different behavior is found. 
 In this case the photoassociation rate decreases with increasing trap 
 frequency, as can be seen especially from the insert of 
 Fig.\,\ref{fig:PAtransmomAttract}(a). In fact, a sharp cut-off of the 
 transition rate is observed. The transitions to the states that 
 possessed the largest photoassociation rate for small trap frequencies 
 are almost completely suppressed for large trap frequencies. Clearly, 
 the simple assumption ``a tighter trap leads to a higher photoassociation 
 rate due to an increased spatial confinement'' is only partly true. 
 The fact that this assumption cannot be valid for 
 all final states can be substantiated by means of a general sum-rule  
 that is derived and discussed in the following subsection.

%--------------------------------------------------------------------------

\subsection{Sum rule}
\label{sec:sumrule}

 Performing a summation (including for $\omega=0$ an integration over the
 dissociative continuum) over all final vibrational states (using closure) 
 yields
\begin{equation}               
\ds
  \widetilde{I}(\omega) \; =\; 
       \sum\limits_{v=0}^\infty \: I^v(\omega) \;=\; 
       \int\limits_0^\infty\: \Psi^{\rm 10'}(R;\omega) \, D^2 (R) \, 
                     \Psi^{\rm 10'}(R;\omega) \:{\rm d}R \; .
\label{sumrule}
\end{equation}
 While the electronic transition dipole moment $D(R)$ depends clearly  
 on $R$ for small internuclear separations, it reaches its asymptotic 
 value (the sum of the electronic dipole transition moments of two 
 separated atoms, $D_{\rm at}$) at some $R$ value that is much 
 smaller than the typical spatial extend of the final vibrational 
 states with the largest transition amplitudes.  
 (In the example of Li$_2$ this asymptotic limit is reached at 
 about $25\, a_0$.) If the largest photoassociation amplitudes  
 result from transitions to final states whose wave functions 
 are mostly located outside this $R$ 
 range, the integral in Eq.\,(\ref{sumrule}) is dominated by the $R$ 
 regime in which $D(R)$ is constant. In this case $D^2$ can be taken out 
 of the integral and normalization of the initial wave function assures
 $\widetilde{I}(\omega) \approx \widetilde{I} = D_{\rm at}^2$. 
 
 Consequently, for all trap frequencies that are too small to confine
 the atoms into a spatial volume that is comparable to the atomic volumes 
 (leading to $D(R)\neq \,D_{\rm at}$) and thus for all traps relevant 
 to this work (and presently experimentally achievable) the total
 dipole transition moment $\widetilde{I}$ is to a good approximation 
 independent of the trap frequency $\omega$. 
 Therefore, changing the trap frequency can only redistribute transition 
 probabilities between different final vibrational states. Increasing the 
 transition rate to one final state must be compensated by a decrease of 
 the transition probability to one or more other vibrational states. 

 A conservative estimate of 
 the minimum and maximum influence of a harmonic trap on the photoassociation 
 rate is obtained from $\widetilde{I}_{\rm min}=D_{\rm min}^2$ and 
 $\widetilde{I}_{\rm max}=D_{\rm max}^2$, respectively, where $D_{\rm min}$ 
 ($D_{\rm max}$) is the minimum (maximum) value of the molecular electronic 
 transition dipole moment.  

 The sum-rule values obtained numerically for the trap 
 frequencies shown in Fig.\,\ref{fig:PAtransmomAttract}(a) are
 $\widetilde{I}(\omega = 2\pi\times 1\,$kHz$) = 11.127222$,
 $\widetilde{I}(\omega = 2\pi\times 10\,$kHz$) = 11.12723$,
 and $\widetilde{I}(\omega = 2\pi\times 100$\,kHz$) = 11.1273$.
 This may be compared to the value
 $ \lim_{R\rightarrow\infty} \, |D(R)|^2 
  = D_{\rm at}^2\ = |D_{\rm 2s-2s}+D_{\rm 2s-2p}|^2
  = |D_{\rm 2s-2p}|^2
  = 11.1272213$ obtained from the calculation described in 
 Sec.\,~\ref{sec:system}.
 Clearly, the sum-rule~(\ref{sumrule}) can also be used as 
 a test for the correctness of numerical calculations. The 
 very small deviations from the predicted sum-rule value 
 may, however, not only be a result of an inaccuracy of the 
 present numerical approach, but also reflect the (small) 
 $R$ dependence of $D(R)$ that allows some $\omega$
 dependence of the total photoassociation rate. This
 interpretation is supported by the fact that the numerically
 found deviations monotonously increase with increasing frequency
 $\omega$. Larger values of $\omega$ lead to a spatially more
 confined $\Psi^{\rm 10'}(R;\omega)$ which in turn
 probes more of the $R$-dependent part of $D(R)$. Since
 $D(R)$ approaches $D(R=\infty)=D_{\rm at}$ from above,
 a small increase of $\widetilde{I}$ is expected for increasing
 trap frequencies. As is evident from
 Fig.\,\ref{fig:PAtransmomAttract}\,(a) (especially the insert),
 the sum-rule fulfillment is achieved by a drastic decrease of the
 photoassociation rate to the highest lying final states. This 
 compensates the trap-induced increased rate to the lower lying 
 states. Since the rate to the highest lying states is orders of 
 magnitude larger than the one to the low-lying states, the reduced
 transition probability of a small number of states can easily
 compensate the substantial increase by orders of magnitude observed 
 for the large number of low-lying states.

 From Eq.~(\ref{sumrule}) it is clear that in those cases where 
 most of the contribution to the sum rule stems from the $R$ range 
 where $D(R)$ is practically constant, there is also no influence 
 of the initial-state wave function. Taking $D$ out of the integral 
 yields always the self-overlap of the initial-state wave function 
 and thus unity. This is important, since it indicates that the 
 sum-rule value is also (approximately) independent of the 
 atom-atom interaction potential.  

%--------------------------------------------------------------------------

\subsection{Enhancement and suppression factor $f^v$}
\label{sec:ratio}

 In order to quantify the effect of a tight harmonic trap on 
 the photoassociation rate and to get rid of its variation 
 as a function of the final-state vibrational level 
 $v$ (that is due to the nodal structure and clearly visible 
 in Fig.\,\ref{fig:PAtransmomAttract}(a) especially for smaller $v$) 
 the ratio  
\begin{equation}   
\ds
       f^{v}(\omega) = \frac{I^{v}(\omega)}{I^{v}(\omega_{\rm ref})}\; .
\label{f_v}
\end{equation}
 may be introduced. It describes the relative enhancement 
 ($f^{v}(\omega)>1$) or suppression ($f^{v}(\omega)<1$) of 
 the photoassociation rate to a specific final state $v$ 
 at a given trap frequency $\omega$ with respect to the reference 
 frequency~$\omega_{\rm ref}$. Although it may appear to be most 
 natural to choose the trap-free case as reference ($\omega_{\rm ref}=0$), 
 a finite value offers some advantages. First, a different 
 normalization applies to $\omega=0$ and $\omega\neq 0$, 
 since in one case it is free-to-bound transitions,  
 while it is bound-to-bound transitions otherwise.  
 Second, from a numerical point of view it is more convenient 
 to treat both cases the same way and to avoid the variation of the 
 box boundary $\ds R_{\rm max}$ that would otherwise be necessary for 
 the trap-free case. Finally, it may be argued that a non-zero trap reference 
 state is in fact more relevant to typical photoassociation experiments with 
 ultracold alkali atoms, since most of them are anyhow performed in traps. In 
 the present work $\omega_{\rm ref} = 2 \pi \times 1\,\mbox{kHz}$ was 
 chosen. This value is sufficiently small to represent typical shallow 
 traps in which the influence of the trap on photoassociation is supposedly 
 (at least to a good approximation) negligible. On the other hand, it allows 
 to calculate the transition dipole moments with reasonable numerical efforts 
 and thus sufficient accuracy. 

 The ratio $f^{v}(\omega)$ is shown for two different trap frequencies 
 $\omega$ in Fig.\,\ref{fig:PAtransmomAttract}(b). For most of the 
 vibrational final states a simple constant regime is observed, 
 i.\,e.\ the ratio $f^{v}(\omega)$ is independent of $v$ for all except 
 the highest lying states. This constant regime is followed by a relatively 
 sharp cut-off beyond which the ratio $f^{v}(\omega)$ is very small. 
 In the constant regime a 100\,kHz trap leads to an enhancement by 
 almost 3 orders of magnitude. 

 Comparing the results for different $\omega$ one notices that in the  
 range of final states where a constant behavior (with respect to $v$) is 
 observed, a tighter trap leads to an increased photoassociation rate, 
 trap-induced enhanced photoassociation (EPA).  
 Due to the cut-off this is, however, not true, if the last 
 vibrational states are considered. Since the range of constant behavior 
 shrinks with increasing trap frequency, there is an increasing range 
 of vibrational states for which a tighter trap leads to a smaller 
 photoassociation rate compared to a shallower trap. In this case 
 trap-induced suppressed photoassociation (SPA) occurs.
 This effect is especially visible from the insert of 
 Fig.\,\ref{fig:PAtransmomAttract}(a). The physical origin of the two
 different regimes (constant vs.\ cut-off) and their $\omega$ dependence 
 is discussed separately in the following two subsections.

%--------------------------------------------------------------------------

\subsection{Constant regime}
\label{sec:constregim}

 Since in the constant regime the ratio $f^v(\omega)$ is completely
 independent of the final state level $v$, its value (for a given $\omega$) 
 and constancy (as a function of $v$) must be a consequence of the influence 
 of the trap on the initial state. The initial-state wave function for 
 a $^6$Li atom pair was shown for three different trap frequencies in 
 Fig.~\ref{fig:iniWF4wregimes}. A view on the complete wave function 
 reveals directly the confinement of the wave function to a smaller 
 spatial volume, if the trap frequency is increased. However, on this 
 scale the variation of the wave function at a specific value of $R$ 
 appears to be quite complicated. Thus it is not at all clear why the 
 enhancement factor $f^v$ has for so many states a constant value. 
 A closer look at smaller internuclear separations (insert of 
 Fig.~\ref{fig:iniWF4wregimes}) reveals that besides the initial 
 oscillatory part confined to the effective range of the atom-atom 
 interaction potential there is a relatively large $R$ interval in which the 
 wave functions for the different trap frequencies vary linearly with $R$.  
 In this case the slope is very small and the wave function is thus almost 
 constant. If the Franck-Condon integral is determined only by the value of 
 the initial-state wave function in this $R$ window, it produces an almost 
 undistorted image of the final-state wave function. 

\begin{figure}[ht]                    
  \centering
  \includegraphics[width=8.5cm,height=6.5cm]{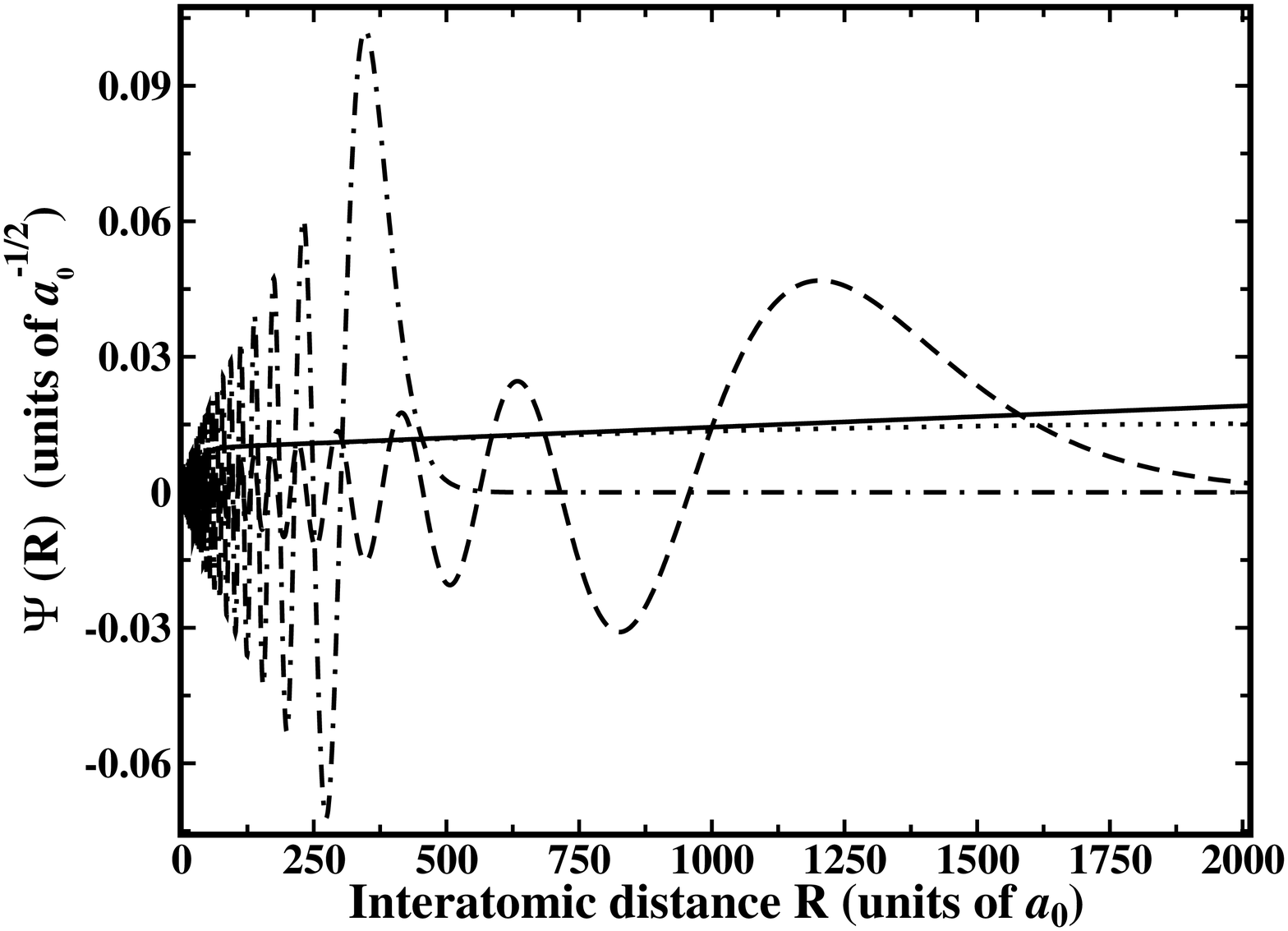}
  \caption{{\footnotesize
      Illustration of the constant photoassociation regime. The 
      initial-state wave function 
      $\Psi^{\rm 10'}(R;\omega=100)$ (dots) and an approximation 
      to it, $\sqrt{f_c(\omega=100)}\cdot\Psi^{\rm 10'}(R;\omega=1)$ 
      (solid), are shown together with the final-state vibrational 
      wave functions for $\ds v= 88$ (chain)
      and $\ds v=94$ (dashes).
    } }\label{fig:constnonconst}
\end{figure}

 However, for the ratio $f^v(\omega)$ this final-state dependence 
 disappears. The reason is that in the $R$ range where the initial-state 
 wave function is almost constant, its variation with the trap frequency 
 is also $R$ independent, as can be seen from the insert of 
 Fig.~\ref{fig:iniWF4wregimes}. In other words, one finds  
 $\ds \Psi^{\rm 10'}(R;\omega)=C(\omega) 
                                 \cdot \Psi^{\rm 10'}(R;\omega_{\rm ref})$. 
 If no final-state dependence occurs, the constant $C(\omega)$ is related
 to the ratio $f$ via $f_c(\omega)=C^2(\omega)$. The validity of this  
 argument for the occurrence of a constant ratio $f$ can thus be checked (and 
 visualized) in the following way. Together with the correct wave function 
 $\ds \Psi^{\rm 10'}(R;\omega)$ the approximate one, 
 $~\sqrt{f_c(\omega)}~\cdot~\Psi^{\rm 10'} (R;\omega_{\rm ref})$, 
 is plotted where $f_c(\omega)$ is the value of the factor $f$ in the
 constant regime. A convenient way to determine $f_c(\omega)$ follows from 
 the observation that the constant regime always starts at $v=0$. Thus 
 $f_c(\omega)=f^{v=0}(\omega)$ is the most straightforward way of 
 $f_c(\omega)$ determination. In Fig.~\ref{fig:constnonconst} the 
 correct wave function $\ds \Psi^{\rm 10'}(R;\omega)$ is plotted 
 together with the approximate wave function 
 $~\sqrt{f_c(\omega)}~\cdot~\Psi^{\rm 10'} (R;\omega_{\rm ref})$  
 for the trap frequency $\ds \omega =2\pi\times\mbox{100kHz}$. 
 The agreement between the two wave functions is clearly good in the 
 shown range of $R$ values, but it is better for small $R$,  
 since at about $R=500\,a_0$ the two wave functions start to disagree. 
 Below $R=500\,a_0$ the two wave functions agree completely with each 
 other, even in the very short $R$ range where they possess an oscillatory 
 behaviour. 

 The key for understanding the occurrence of the constant 
 regime is that a variation of the trap frequency modifies the spatial 
 extent of the initial-state wave function, but leaves its norm and 
 nodal structure preserved. As a consequence, the wave function changes 
 qualitatively only in the large $R$ range, while in the short range 
 only the amplitude varies (by factor $C(\omega)$). The reason for this 
 behaviour is that at small $R$ the wave function is practically shielded 
 from the trap potential by the dominant atom-atom interaction. 

 Fig.~\ref{fig:constnonconst} shows also two final-state wave functions 
 ($v=88$ and 94). According to 
 Fig.~\ref{fig:PAtransmomAttract}\,(b) the transition to $v=88$ belongs 
 still to the constant regime ($f^{88}(\omega)\approx f_c(\omega)$), 
 though to its very end. The transition to $v=94$ does on the other 
 hand not belong to this regime, since for the considered trap 
 frequency $f^{94}(\omega)<f_c(\omega)$. As is evident from 
 Fig.~\ref{fig:constnonconst}, a constant ratio $f^v(\omega)$ 
 is observed as long as the final-state wave function $v$ is completely 
 confined within an $R$ range in which the approximation     
 $\ds \Psi^{\rm 10'}(R;\omega)\approx 
             C(\omega) \cdot \Psi^{\rm 10'}(R;\omega_{\rm ref})$  
 is well fulfilled. This is (for 
 $\omega=2\pi\times\mbox{100kHz}$) the case for $v=88$ for which the
 wave function is confined within $R<600\,a_0$, but not for $v=94$ whose 
 outermost lobe has its maximum at about $1350\,a_0$. Since the 
 $R$ range in which the initial-state wave function can be approximated 
 in the here discussed fashion decreases with increasing trap frequency, 
 the range of $v$ values for which $f^v(\omega)\approx f_c(\omega)$ 
 is valid diminishes with increasing trap frequency. 

 The following rule of thumb is found to determine those vibrational 
 levels $v$ for which the relation $f^v(\omega)\approx f_c(\omega)$ 
 starts to break down. For trap frequencies 
 $\ds \omega_1$ and $\ds \omega_2$ (with $\ds \omega_2 > \omega_1$) 
 one may define a difference $\Delta$ that quantifies the deviation 
 of $C\cdot\Psi^{\rm 10'}(R;\omega_{\rm 1})$
 and $\Psi^{\rm 10'}~(R;\omega_{\rm 2})$ as
 $\Delta (R)~ =~C\cdot~\Psi^{\rm 10'}~(R;\omega_{\rm 1})
 ~ -~\Psi^{\rm 10'}~(R;\omega_{\rm 2})$ where
 $\ds C=\sqrt{f_c(\omega_2)}$. For example, in Fig.~\ref{fig:constnonconst} 
 the difference $\Delta (R)$ is the distance   
 between the solid curve and the dotted one. The relation 
 $f^v(\omega)\approx f_c(\omega)$ breaks down for those final states 
 $v$ whose classical turning point lies beyond $R_0$. $R_0$ itself is 
 determined by $\Delta (R>R_0)\gtrsim 10^{-3}$. In other words, if the 
 last lobe of the final wave function overlaps substantially with a 
 region where the deviation defined by $\Delta$ is larger than 
 about $10^{-3}$, a clear deviation from the constant regime is to 
 be expected.

%--------------------------------------------------------------------------

\subsection{Cut-off regime}
\label{sec:cutoffregim}

\begin{figure}[ht]     
 \centering
\includegraphics[width=8.5cm,height=6.5cm]{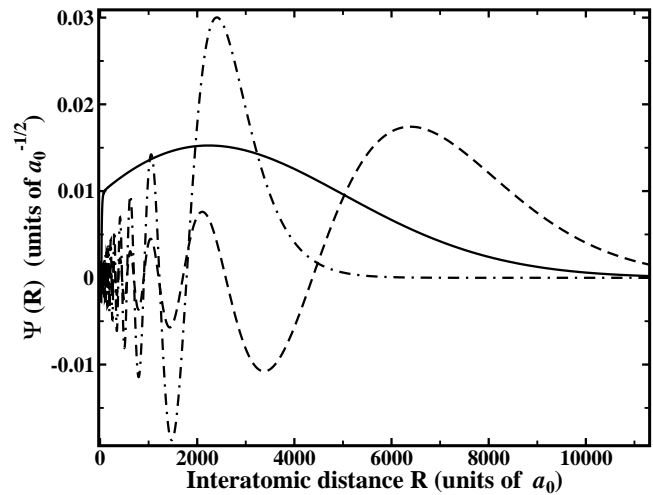}
 \caption{{\footnotesize
     Illustration of the cut-off regime for a 100\,kHz trap. 
     The initial-state wave function $\ds \Psi^{\rm 10'}(R;100)$ (solid) 
     is shown together with the two final-state wavefunctions for 
     $v=96$ (chain) and $v=98$ (dashes).  
} }\label{fig:cutoff}
 \end{figure}

 Once the constant regime of the ratio $f^v$ (for a given trap frequency) is
 left, $f^v$ is steadily decreasing with $v$, as is apparent from  
 Fig.~\ref{fig:PAtransmomAttract}\,(b). The photoassociation rate displays 
 then a rather sharp cut-off behavior 
 (see insert of Fig.~\ref{fig:PAtransmomAttract}\,(a)). The most loosely bound
 vibrational states of the final electronic state have in the trap-free 
 case the largest rate but possess a very small one in very tight 
 traps. For those high-lying states the wave functions have a very highly 
 oscillatory behavior for short $R$ values and a large lobe close to the 
 classical turning point. This outermost lobe determines the Franck-Condon 
 integral, if the initial-state wave function is sufficiently smooth in this 
 $R$ range. In Fig.~\ref{fig:cutoff}
 the initial-state wave function is shown together 
 with the ones for $v=96$ and 98 (for 
 $\ds \omega = 2\pi\times\mbox{100kHz}$).

 It is evident from Fig.~\ref{fig:cutoff} that for 
 $\ds v=96$ the overlap of the initial wave function with the last lobe 
 of the final state is very large. In fact, for this trap frequency 
 the overlap reaches its maximum for $v=96$ and $97$ (see 
 Fig.~\ref{fig:PAtransmomAttract}\,(a)), despite the fact that the 
 trap-induced relative enhancement factor $f^v(\omega)$ is small 
 (Fig.~\ref{fig:PAtransmomAttract}\,(b)). In the case of $v=98$ the 
 transition rate is not only clearly smaller than for $v=96$ or $97$, 
 but it is also much smaller than the rate obtained for the same level  
 at much lower trap frequencies (10 or 1\,kHz). Clearly, one has 
 $f^{v=98} (\omega)<1$ and thus for  
 $\ds \omega = 2\pi\times\mbox{100kHz}$ the level $v=98$ represents 
 an example for a trap-induced suppressed rate (SPA) in contrast to 
 the usually expected enhanced photoassociation in a trap (EPA, 
 $f^v(\omega)>1$).
 From Fig.~\ref{fig:cutoff} it is clear that the 
 reason for the small transition rate to $v=98$ is due to the fact that 
 the outermost lobe of the $v=98$ state lies mostly outside 
 the $R$ range in which the initial-state wave function is non-zero. The 
 least bound state (in the trap free case), $v=99$, possesses an even 
 smaller photoassociation rate, since in this case the outermost lobe lies 
 practically completely outside the non-zero $R$ range of the initial-state 
 wave function. Due to the imperfect cancellation of the oscillating 
 contributions from the inner lobes, the photoassociation rate 
 for $v=99$ is very small, but non-zero. 

 Increasing the trap frequency even more will confine the initial-state 
 wave function to a smaller $R$ range and thus SPA occurs for smaller 
 $v$ values. The origin of the suppression is in fact a quite remarkable 
 feature, since from Fig.~\ref{fig:cutoff} it is clear that the trap has 
 practically no influence on the final states, even if one considers the 
 highest-lying ones that have very tiny binding energies. This is still  
 true, if the spatial extent of the final state is much larger than the 
 one of the trap potential. This may be interpreted as a shielding 
 of the trap potential by the molecular (atom-atom interaction) 
 potential. The reason for the different shielding experienced by  
 the initial and the final states is not only due to the fact that 
 the former lies above the dissociation threshold, since then the 
 photoassociation rate should dramatically increase, if transitions 
 into the purely trap-induced bound states of the final electronic 
 state are considered. This is, however, not the case as can be seen 
 for the states $v>99$ in Fig.~\ref{fig:PAtransmomAttract}\,(a). 
 The different shielding is due to the inherently different long-range 
 behaviors of the two electronic potential curves describing the initial 
 ($a ^3\Sigma^+_u$) and the final ($1 ^3\Sigma^+_g$) state. 
 If one introduces the crossing point $\ds R_c$ of the long range part 
 of the van der Waals potential with the one of the inverted harmonic  
 trap, it is defined by equating $\ds C_n/R_c^n$ and 
 $\ds \frac12\mu\omega^2R_c^2$ where $C_n$ is the corresponding leading 
 van der Waals coefficient. At the point $\ds R_c$ the trap  
 potential starts to dominate. For example, in the case of the trap 
 frequency $\omega = 2\pi\times\mbox{10kHz}$ one finds 
 $\ds R_c \approx +825\, a_0$  and $\ds R_c \approx +17700\, a_0$ 
 for $a ^3\Sigma^+_u$ and $1 ^3\Sigma^+_g$ of Li$_2$, respectively.

%--------------------------------------------------------------------------

\subsection{$\ds I^{v}(\omega)$ for a repulsive interaction}
\label{sec:positivesclen}

 In order to check the main conclusions of the results obtained 
 for $^6$Li$_2$ also the formation of $^{39}$K$_2$ is investigated.
 While for $^6$Li a photoassociation process
 between {\it triplet} states was considered, a transition between 
 the $X ^1\Sigma^+_g$ and the $A ^1\Sigma_u^+$ states is chosen  
 for $^{39}$K. In contrast to the large negative scattering length 
 of two $ ^6$Li atoms interacting via the $a ^3\Sigma^+_u$ potential 
 two $^{39}$K ground-state atoms interact via a small positive s-wave 
 scattering length. The obtained results for the squared transition 
 dipole moments $I^v(\omega)$ are qualitatively very similar to the 
 results obtained for $^6$Li$_2$. This includes the existence of a 
 constant regime of $f^v(\omega)$ followed by a pronounced decrease 
 for the highest-lying vibrational states, the cut-off. The rule of 
 thumb for predicting the range of $v$ values for which a constant ratio 
 $f^v$ is observed does also work in this case.   
 $^{39}$K$_2$ shows thus trap-induced suppressed photoassociation 
 for the highest lying states with a sharp cut-off in the 
 $I^v(\omega)$ spectrum very much like $^6$Li$_2$. Therefore, 
 the results are not explicitly shown for space reasons.

\begin{figure}[ht]           
 \centering
\includegraphics[width=8.5cm,height=13cm]{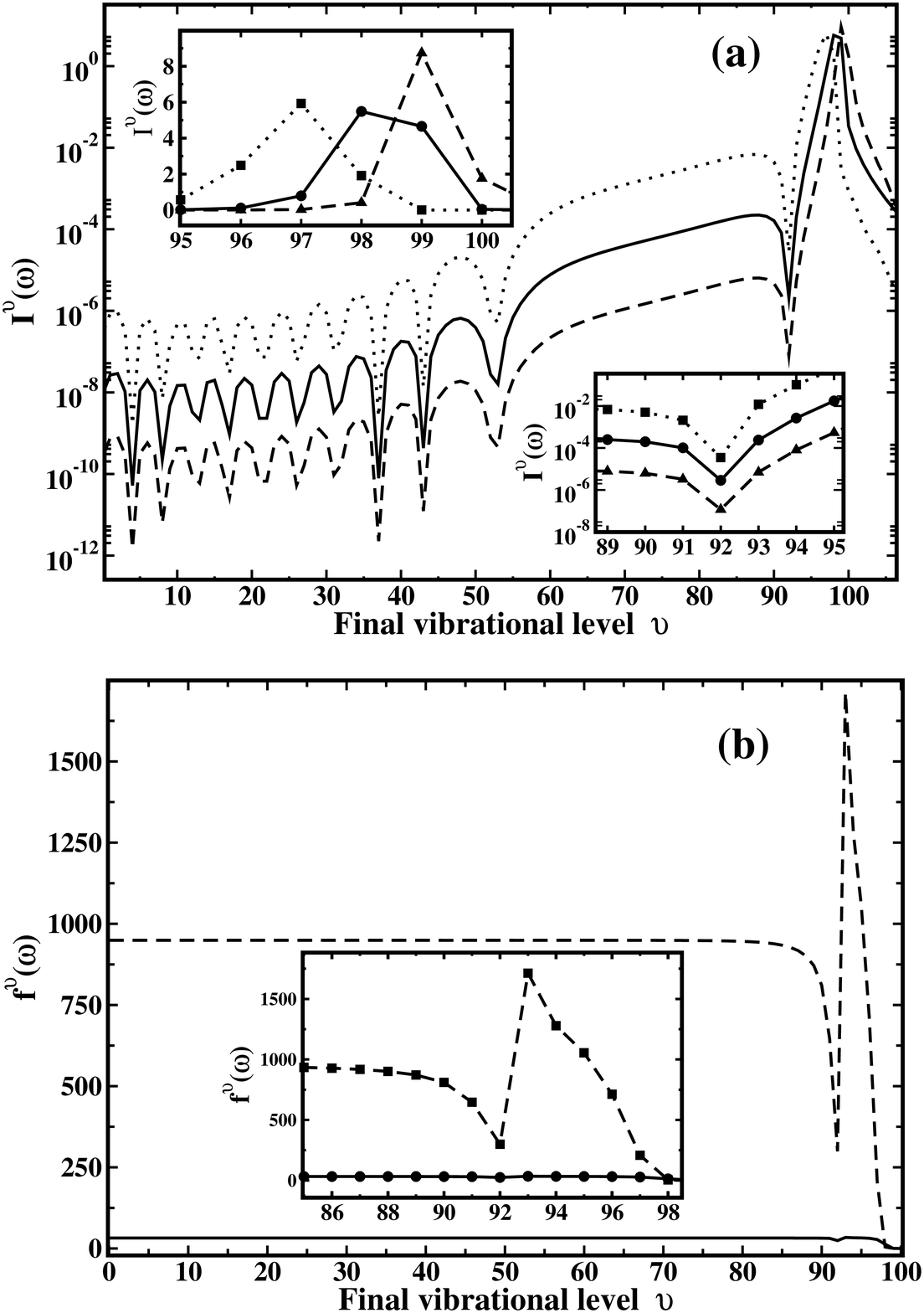}
 \caption{{\footnotesize
     As Fig.~\ref{fig:PAtransmomAttract} 
     but the scattering length is artificially changed to 
     $a_{\rm sc}=+850\,a_0$ (see text for details).  
     The additional insert in the right bottom corner of (a) 
     shows the range $v=89$ to $v=95$ on an enlarged scale. 
}}\label{fig:PAtransmomRepuls}
 \end{figure}

 For a more systematic investigation of the influence of the 
 scattering length $a_{\rm sc}$ and thus the type of interaction (sign 
 of $a_{\rm sc}$) and its strength (absolute value of $a_{\rm sc}$) 
 the mass of the Li atoms is varied. The mass variation allows 
 for an in principle continuous (though non-physical) modification 
 of $a_{\rm sc}$ from very large positive to negative values. With 
 increasing mass an increasing number of bound states is supported 
 by the same potential curve.  Since $a_{\rm sc}$ is sensitive to the 
 position of the least bound state, even a very small mass variation 
 has a very large effect, if a formerly unbound state becomes 
 bound. For example, an increase of the mass of $ ^6$Li by 0.3\% 
 changes $a_{\rm sc}$ from $-2030\,a_0$ to about $+850\, a_0$. 
 The (for $^6$Li unbound) 11th vibrational state becomes weakly bound. 
 A further increase of the mass increases its binding energy until  
 it reaches the value for $^7$Li. It is also possible to modify 
 $a_{\rm sc}$ from $-2030\,a_0$ to $+850\, a_0$ by lowering the mass 
 of $^6$Li. A larger mass variation is required (about 
 18\,\%) but the number of bound states remains unchanged. In this 
 case the large positive value of $a_{\rm sc}$ indicates that the 
 10th bound state is, however, only very weakly bound and a further 
 small decrease of the mass will shift it into the dissociative 
 continuum. 

\begin{figure}[ht]     
 \centering
\includegraphics[width=9cm,height=6cm]{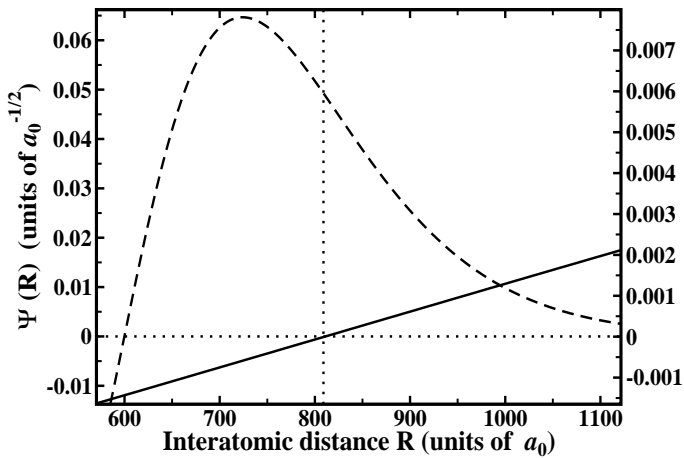}
 \caption{{\footnotesize
     For $^6$Li$_2$ (the scattering length in the initial channel is 
     modified to $a_{sc} = +850\, a_0$) and for a 100\,kHz trap 
     the outermost lobe of the final-state wave function ($v=92$, long 
     dashes, left scale) is shown together with the initial-state wave 
     function ($v'=10$, solid, right scale). 
     Auxiliary horizontal and vertical dotted lines are given to 
     assist the eye in identifying regions with a positive or negative   
     overlap of the wavefunctions.
} }\label{fig:dip}
 \end{figure}

 In Fig.~\ref{fig:PAtransmomRepuls}\,(a) $I^v(\omega)$ is shown 
 for $a_{\rm sc}=+850\,a_0$ (achieved by a 0.3\% increase of the 
 mass) and three different trap frequencies as an example for a 
 large positive scattering length and thus strong repulsive 
 interaction. The overall result is again very similar to the one 
 obtained for a large negative scattering length. A tighter trap 
 increases the transition rate for most of the states, but there is a 
 sharp cut-off for large $v$. The position of this cut-off 
 moves to smaller $v$ as the trap frequency is increased. 
 However, for a large positive value of $a_{\rm sc}$ an 
 additional feature appears in the transition spectrum: a 
 photoassociation window visible as a pronounced dip in 
 the $I^v$ spectrum for large $v$. For the given choice of 
 $a_{\rm sc}$ this minimum occurs for $v=92$. 

 The occurence of the dip for $a_{\rm sc}\gg 0$ has been 
 predicted and explained for the trap-free case 
 in~\cite{cold:cote95,cold:cote98b} and was experimentally 
 confirmed~\cite{cold:abra96}. Fig.~\ref{fig:dip} shows the last lobe of 
 the final-state vibrational wave function $\Psi^{\rm 92}(R)$ 
 together with the initial-state wave function, both for 
 $\omega=2\pi\times 100\,$kHz. The key for understanding 
 the occurrence of the dip for large positive scattering 
 lengths and its absence for negative ones is the change 
 of sign of the initial-state wave function as a consequence 
 of the repulsive atom-atom interaction. In fact, in the 
 trap-free case the position of this node agrees of course 
 with the scattering length. As can be seen 
 from Fig.~\ref{fig:dip}, the tight trap moves the nodal position 
 to a smaller value, but this shift is comparatively small 
 (about 5\,\%) even in the case of a 100\,kHz trap. For negative values 
 of $a_{\rm sc}$ this node appears to be absent, since in this 
 case only the extrapolated wave function intersects the $R$ axis, 
 but this occurs at the non-physical interatomic 
 separation $R_{\rm x}=-a_{\rm sc}$. As a result of the sign change 
 occurring for $a_{\rm sc}>0$ the overlap of the 
 initial-state wave function with a final state for which the 
 mean position of the outermost lobe agrees with the nodal 
 position ($R_{\rm x}$) vanishes. The probability for a perfect 
 agreement of those two positions is of course rather unlikely, but 
 as can be seen from  Fig.~\ref{fig:PAtransmomRepuls}\,(a) 
 and~\cite{cold:cote98b} where also an approximation for $I^v(\omega=0)$ 
 was derived, the cancellation can be very efficient.  

\begin{figure}[ht]     
 \centering
\includegraphics[width=8.5cm,height=6.5cm]{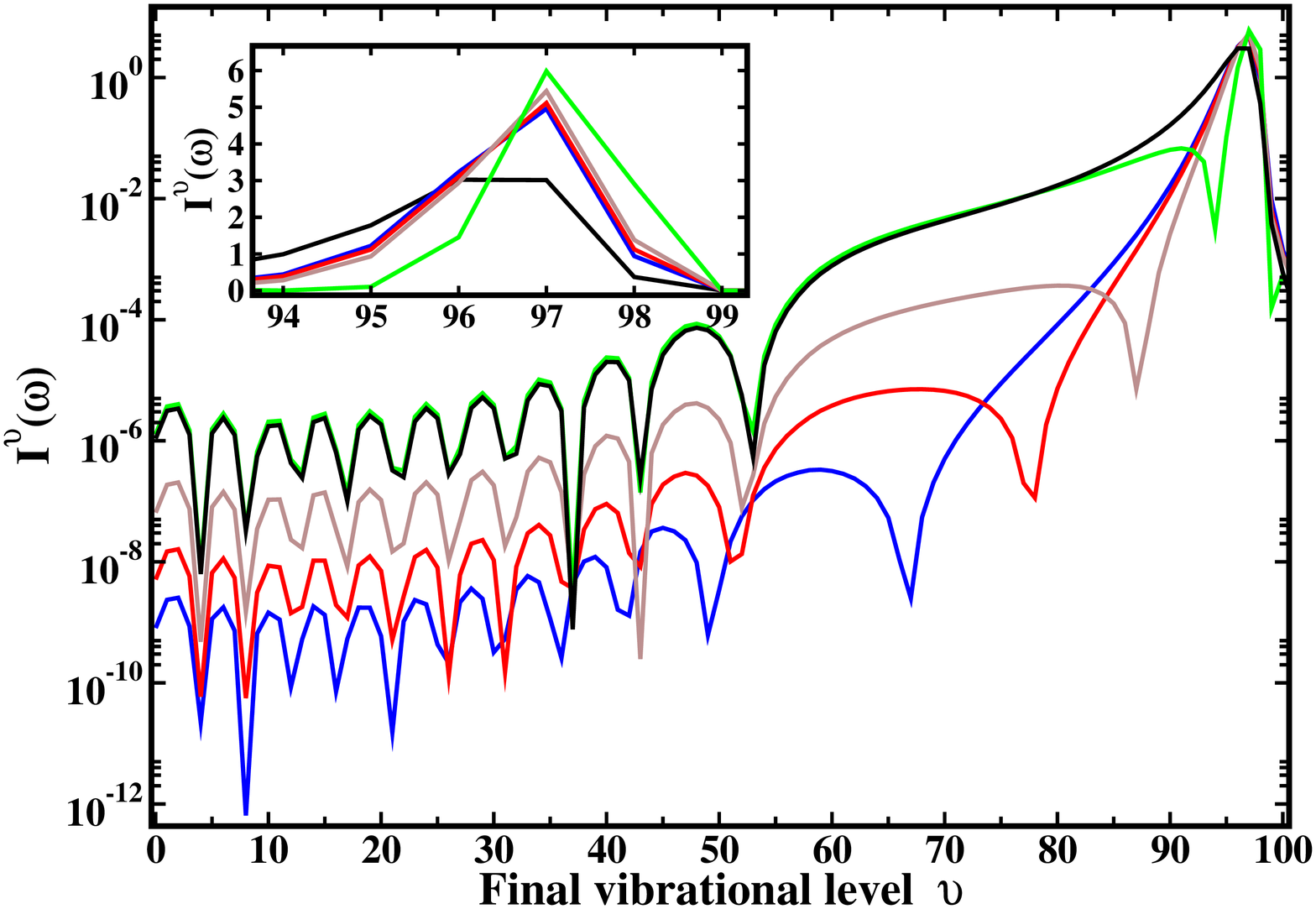}
 \caption{{\footnotesize (Color online) 
         Dependence of the squared dipole transition moments 
         $\ds I^{v}(\omega)$ on the scattering length $a_{\rm sc}$ 
         for transitions from the first trap-induced ($v'=11$) initial 
         $a ^3\Sigma^+_u$ state to the vibrational manifold ($v$) of 
         the $1 ^3\Sigma^+_g$ state of $\mbox{Li}_2$ 
         in a $\ds \omega = 2\pi\times\mbox{100\,kHz}$ trap. 
         Using masses slightly larger than the one of $^6$Li 
         $\ds a_{sc}= +50\, a_0$ (blue), $\ds a_{sc}= +115\, a_0$ (red), 
         $\ds a_{sc}= +350\, a_0$ (brown), and 
         $\ds a_{sc}= +2020\, a_0$ (green) were yielded. For comparison, 
         the result with the physical mass ($\ds a_{sc}=-2030\,a_0$, black) 
         is also shown. The insert shows the transitions to $v=94$ to 99 
         on an enlarged scale.
}}\label{fig:PArateascvarDifAscN11}
 \end{figure}
 It should be emphasised that of course also for $a_{\rm sc}<0$ a 
 number of dips occur as was discussed in the context of 
 Fig.\,\ref{fig:PAtransmomAttract}. The difference between those 
 dips and the one discussed for $a_{\rm sc}\gg 0$ is the 
 occurrence of the latter outside the molecular regime. While the 
 other dips are a direct consequence of the short-range part of the 
 atom-atom interaction potential and thus confined (for Li$_2$) 
 to $v<55$ corresponding to $R<30\,a_0$, the dip occuring for 
 $a_{\rm sc}\gg 0$ can be located outside the molecular regime.
 This is even more apparent from Fig.\,\ref{fig:PArateascvarDifAscN11} 
 where the $I^v$ spectra for four different positive values of $a_{\rm sc}$ 
 are shown together with the one for the (physical) 
 value $a_{\rm sc}=-2030\,a_0$ (all for $\omega=2\,\pi\times 100\,$kHz). 
 The values 
 $a_{\rm sc}=+2020\,a_0$, $+350\,a_0$, $+115\,a_0$, and $+50\,a_0$ were 
 obtained by a mass increase of $\sim 0.3$\%,$\sim 0.8$\%,$\sim 2$\%, and 
 $\sim 6$\%, respectively. In agreement with the explanation given above, 
 the position of the dip moves continuously to larger values of $v$ as 
 the scattering length increases, since the position $R_{\rm x}$ of the 
 last node of the initial state lies close to $a_{\rm sc}$. Also the 
 positions of the other dips depend on $a_{\rm sc}$, but their 
 dependence is much weaker and involves a much smaller $R$ interval. 
 Clearly, the positions of the dips become more stable if they occur 
 at smaller $v$. 

 Noteworthy, the positions of the first 10 dips agree perfectly for 
 $a_{\rm sc}=-2030$ and $+2020\,a_0$. In fact, both spectra are on a first 
 glance in almost perfect overall agreement, except the occurrence of the 
 additional dip for $v=92$. According to the discussion of the sum rule 
 in Sec.\,\ref{sec:sumrule} the total sum $\tilde{I}$ should be 
 (approximately) independent of the atomic interaction and thus $a_{\rm sc}$. 
 This is also confirmed numerically for the present examples. The insert 
 of Fig.\,\ref{fig:PArateascvarDifAscN11} reveals how the sum-rule 
 is fulfilled. The due to the additional dip missing transition 
 probability is compensated by an enhanced rate to the neighbor states 
 with larger $v$.      

\begin{figure}[ht]     
 \centering
\includegraphics[width=8.5cm,height=6.5cm]{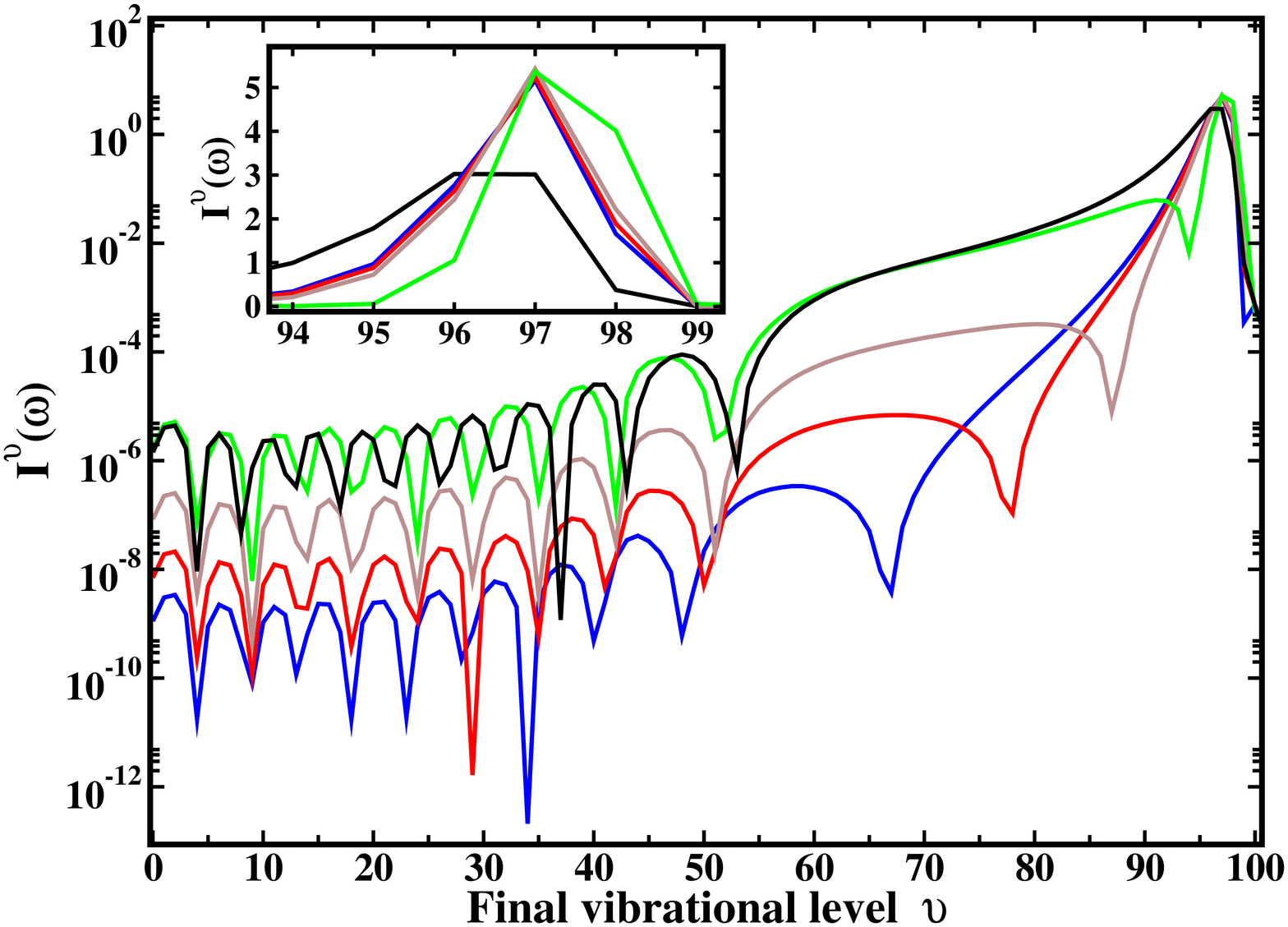}
 \caption{{\footnotesize (Color online) 
     As Fig.~\ref{fig:PArateascvarDifAscN11}, but for a variation of 
     $a_{\rm sc}$ by means of a reduction of the mass with respect to 
     the one of $^6$Li. In this case, the number of bound states remains 
     unchanged and the transition starts from the $v'=10$  
     $a ^3\Sigma^+_u$ state. 
}}\label{fig:PArateascvarDifAscN10}
 \end{figure}
 In all shown cases with $a_{\rm sc}>0$ there exist 11 bound states in 
 contrast to the 10 states of $^6$Li ($a_{\rm sc}=-2030\,a_0$). As 
 mentioned in the beginning of this section, it is also possible to 
 change the sign of $a_{\rm sc}$ while preserving the number of nodes. 
 The corresponding $I^v$ spectra (again for $\omega=2\pi\times 100\,$kHz) 
 are shown in Fig.\,\ref{fig:PArateascvarDifAscN10}. The same values of 
 $a_{\rm sc}$ as in Fig.\,\ref{fig:PArateascvarDifAscN11} ($+2020\,a_0$, 
 $+350\,a_0$, $+115\,a_0$, and $+50\,a_0$) are now 
 obtained by a decrease of the mass by $\sim 18$\%, $\sim 17.5$\%, 
 $\sim 16$\%, and $\sim 13$\%, respectively. A comparison of the two 
 Figs.\,\ref{fig:PArateascvarDifAscN11} and 
 \ref{fig:PArateascvarDifAscN10} demonstrates that the position of the 
 outermost dip (for $a_{\rm sc}\gg 0$) depends for a given $\omega$ solely 
 on $a_{\rm sc}$, while the other dips (in the molecular regime) differ 
 when changing the total number of bound states from 10 to 11. A comparison 
 of the results obtained for $a_{\rm sc}=-2030\,a_0$ and $+2020\,a_0$ 
 with 10 bound states in both cases shows that most of the nodes in the 
 molecular regime are shifted with respect to each other in such a way 
 that the $v$ range hosting 10 dips for $a_{\rm sc}=-2030\,a_0$ contains 
 9 dips for $a_{\rm sc}=-2030\,a_0$.  

 Turning back to Fig.\,\ref{fig:PAtransmomRepuls} and the question of 
 the influence of a tight trap on the photoassociation rate for 
 $a_{\rm sc}\gg 0$ one notices that the position of the additional dip 
 appears to be practically independent of $\omega$. As was explained in 
 the context of Fig.\,\ref{fig:dip}, the reason is that the position of 
 the outermost node depends only weakly on $\omega$. For the shown example 
 this shift is even for a 100\,kHz trap small compared to the separation of 
 the outermost lobes between neighboring $v$ states. Therefore, the shift 
 is not sufficient to move the dip position away from $v=92$. However, 
 if $a_{\rm sc}$ is, e.\,g.\, increased to $+2020\,a_0$ the crossing 
 point $R_{\rm x}$ shifts in a 100\,kHz trap to about $1500\,a_0$ 
 and changes thus by $\approx 25\,\%$. In this case the dip position 
 moves from $v=95$ to $94$. It is therefore important to take the 
 effects of a tight trap into account, if they are used for the 
 determination of $a_{\rm sc}$ using photoassociation spectroscopy  
 the way discussed in~\cite{cold:cote95,cold:abra96}.   

 In order to focus on the effect of the tight trap it is again of 
 interest to consider the ratio $f^v(\omega)$ introduced in 
 Sec.\,\ref{sec:ratio}. For small but positive values of $a_{\rm sc}$ 
 the ratio $f^v$ is structurally very similar to the case 
 $a_{\rm sc}=-2030\,a_0$ shown in Fig.\,\ref{fig:PAtransmomAttract}\,(b). 
 A uniform constant regime covering almost all $v$ states is followed 
 by a sharp cut-off whose position shifts to smaller $v$ as $\omega$ 
 increases. A similar behavior is encountered for $a_{\rm sc}=+850\,a_0$ 
 and $\omega=10\,$kHz as shown in Fig.\,\ref{fig:PAtransmomRepuls}\,(b). 
 However, for a tighter trap (100\,kHz) a new feature appears. In this 
 case the relative enhancement at the dip position ($v=92$) is smaller 
 than in the constant regime, but larger for the neighbor states. 
 The enhancement factor for $v=92$ is only $\approx 25\,\%$ of $f_c$, 
 while the one for $v=93$ is $\approx 60\,\%$ larger than $f_c$. 
 This results in a dispersion-like structure in $f^v$. It should be 
 emphasised that this is again remarkably different from the other dips 
 in $I^v(\omega)$ ($v<55$) that show the same (constant) enhancement 
 factor $f_c$ as their neighbor states.

%--------------------------------------------------------------------------

\subsection{Combined influence of trap and atomic interaction}
\label{sec:combined}

 In view of the very important question how the efficiency of 
 photoassociation can be improved, 
 Fig.\,\ref{fig:PArateascvarDifAscN11} 
 reveals that besides the use of a tight trap a large scattering 
 length is also favorable. The photoassociation rate (away from 
 the dips) is enhanced by orders of magnitude, if $a_{\rm sc}$ 
 varies from $a_{\rm sc}=+50\,a_0$ to $a_{\rm sc}=+2020\,a_0$! 
 In view of the already discussed fact that the results for 
 the overall spectrum $I^v$ differ for $a_{\rm sc}>0$ and 
 $a_{\rm sc}<0$ only by the position of the dips, it is evident 
 that photoassociation (or corresponding Raman transitions) 
 are much more efficient, if $|a_{\rm sc}|$ is very large. 

\begin{figure}[ht]     
\centering
\includegraphics[width=8.5cm,height=6.5cm]{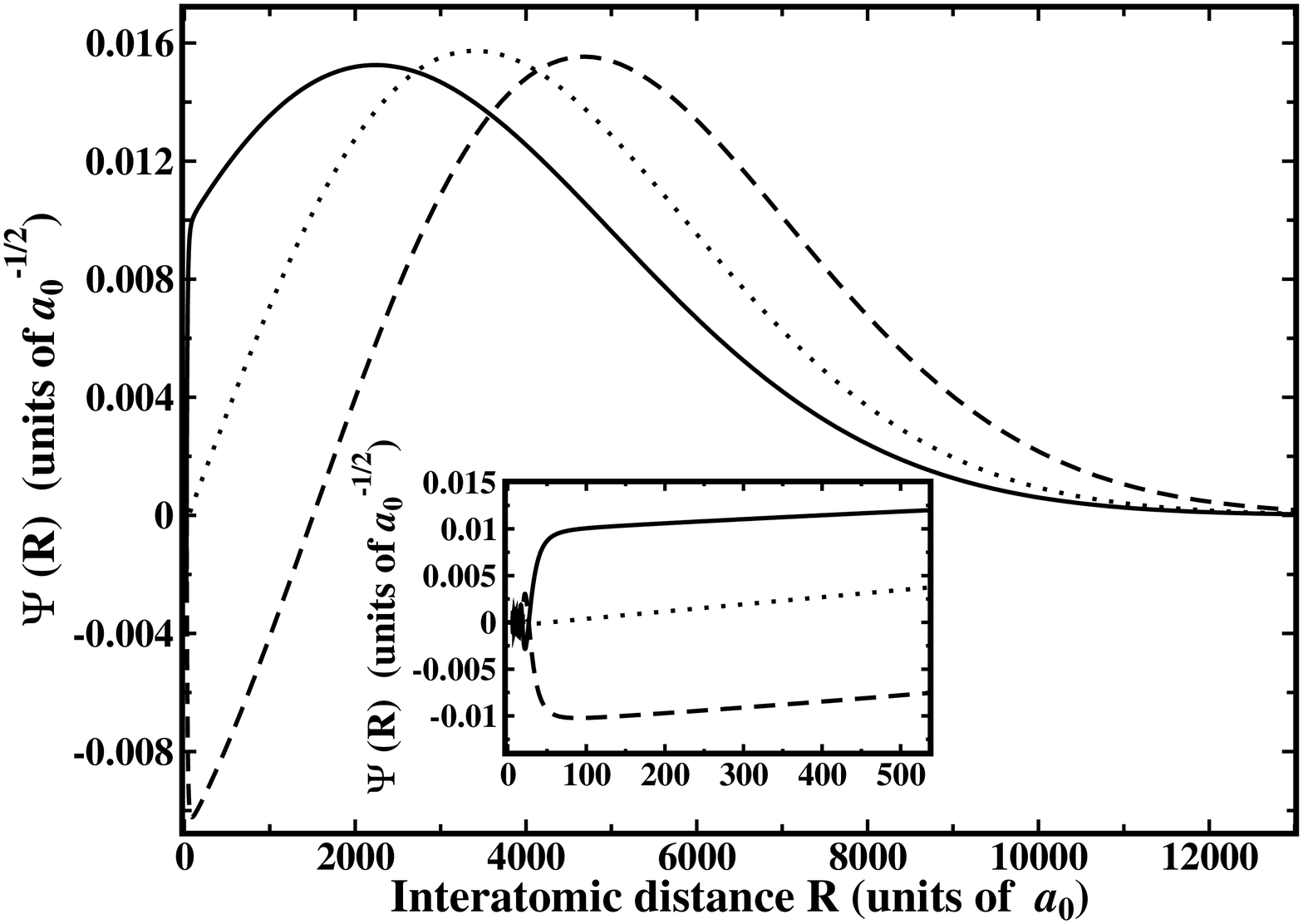}
\caption{{\footnotesize Wave functions of the initial state describing  
two Li atoms in a trap with frequency 
$\ds \omega = 2\pi\times 100\,\mbox{kHz}$ for different masses that 
yield the scattering lengths $a_{\rm sc} = - 2030\,a_0$ 
(solid), $a_{\rm sc} = +50\,a_0$ (dots), and $a_{\rm sc} = +2020\,a_0$ 
(dashes). The insert shows the small $R$ range on an enlarged scale. 
}}\label{fig:AmplitudeDiffer}
\end{figure}

 In order to understand the dependence of the FC factors of the
 vibrational final states on the scattering length it is
 instructive to look at the variation of the initial-state 
 wave function with $a_{\rm sc}$ for large $R$ values. This is 
 shown in Fig.~\ref{fig:AmplitudeDiffer} for 
 $\ds \omega = 2\pi\times\mbox{100\,kHz}$. 
 While a large attractive interaction ($a_{\rm sc}\ll 0$) 
 leads to a very confined wave function for the first trap-induced 
 bound state, a large repulsive interaction ($a_{\rm sc}\gg 0$) 
 does not only result in a node (responsible for the photoassociation 
 window discussed above), but also to a push of the outermost 
 lobe to larger $R$ values. This push is of course counteracted 
 by the confinement of the trap. 
 However, only the highest lying final states probe the very large 
 $R$ range. As is apparent from Fig.~\ref{fig:ClasTurnPoint} the 
 final states $v\le 92$ probe almost completely the range 
 $R\le 1000\,a_0$. Within this $R$ interval the absolute value of 
 the initial-state wave function increases with the absolute value 
 of $a_{\rm sc}$. As a consequence, the corresponding FC factors 
 and $I^v$ should increase with $|a_{\rm sc}|$. An exception to 
 this is the already discussed occurrence of the photoassociation 
 window (spectral dip) that occurs for a positive scattering length, 
 if the position of the node is probed by the final-state wave function.     
 Consequently, one expects for the low-lying final states (in fact 
 for almost all except the very high-lying ones and the ones at the 
 dip position) that an increase of $|a_{\rm sc}|$ leads to an 
 increased photoassociation rate. 
 
 An evident question is of course, whether the enhancements  
 due to the use of tighter traps and tuning of $a_{\rm sc}$ can 
 be used in a constructive fashion? In order to investigate 
 this question, one can introduce another enhancement factor % 
 \begin{equation}  
    g^v(\omega,a_{\rm sc}) \;=\; \frac{I^v(\omega,a_{\rm sc})}
                            {I^v(\omega_{\rm ref},a_{\rm sc,ref})}
  \label{eq:g_v}
 \end{equation}
 with $a_{\rm sc,ref}=0\,a_0$ (and $\omega_{\rm ref}
 =2\pi\times 1\,$kHz as before). Clearly, a cut through 
 $g^v(\omega,a_{\rm sc})$ for constant $a_{\rm sc}$ is 
 equal to $f^v(\omega)$. A cut for constant $\omega$ 
 describes on the other hand the relative enhancement 
 of the photoassociation rate as a function of $a_{\rm sc}$. 
 
 The function $g^v(\omega,a_{\rm sc})$ depends of course on 
 the vibrational state $v$, but as was discussed before, 
 most of the states show a constant enhancement factor $f_c$. 
 Thus it is most interestingly to investigate   
 $g_c(\omega,a_{\rm sc})$ that is defined as the $g$ function  
 for vibrational states for which the relation $f^v = f_c$ is 
 valid. This excludes the states in the cut-off regime and 
 those at or very close to the photoassociation window. 
\begin{figure}[ht]     
 \centering
\includegraphics[width=8.5cm,height=6.5cm]{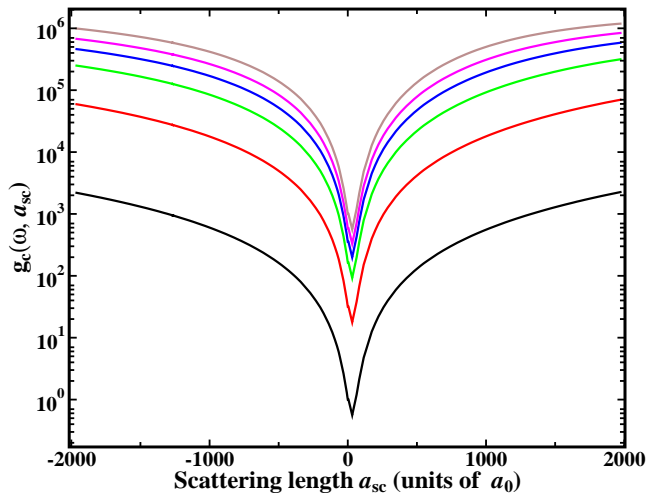}
 \caption{{\footnotesize (Color online) 
     Enhancement factor $g_c$ [see Eq.\,(\ref{eq:g_v})] in the constant 
     ($v$-independent) regime as a function of $a_{\rm sc}$ for   
     trap frequencies $\omega = 2\pi\times 1\,$kHz (black), 
     $\omega = 2\pi\times 10\,$kHz (red),
     $\omega = 2\pi\times 30\,$kHz (green),
     $\omega = 2\pi\times 50\,$kHz (blue),
     $\omega = 2\pi\times 70\,$kHz (purple), 
     and $\omega = 2\pi\times 100\,$kHz (brown). 
}}\label{fig:w1TOw100_ascM1970TOP2020}
 \end{figure}
 In Fig.~\ref{fig:w1TOw100_ascM1970TOP2020} $g_c(\omega,a_{\rm sc})$ 
 is shown as a function of $a_{\rm sc}$ for different trap  
 frequencies. The important finding is that $g_c(\omega,a_{\rm sc})$ 
 increases as a function of $\omega$ and $|a_{\rm sc}|$. In fact, 
 within the shown ranges of $\omega$ and $a_{\rm sc}$ the function 
 $g_c(\omega,a_{\rm sc})$ raises by 6 orders of magnitude, 
 if the maximum values of $\omega$ ($2\pi\times 100\,$kHz) 
 and $a_{\rm sc}$ ($\pm 2000\,a_0$) are considered! 
 A more detailed analysis shows that the enhancement is 
 almost equally distributed among the two parameters, 
 i.\,e.\ a factor 10$^3$ stems from the variation of $\omega$ 
 and about the same factor from varying $a_{\rm sc}$. Thus the 
 enhancement of the photoassociation rate due to the two 
 different physical parameters occurs practically independently  
 of each other, at least in the rather large parameter range 
 considered. It should be emphasized that these ranges  
 are realistically achievable in present-day experiments.  
 It is interesting to note that this finding is not only 
 very encouraging with respect to the possible enhancement of 
 photoassociation rates and related molecule production 
 schemes, but it shows also that the influence of the 
 parameters scattering ($a_{\rm sc}$) and characteristic 
 length scale of an isotropic harmonic trap 
 ($\ds a_{\rm ho} = \sqrt{1/(\mu\omega)}$) on the 
 photoassociation process is very different from the one observed for 
 the energy. In energy-related discussions (like the one on the validity 
 of the pseudopotential approximation in~\cite{cold:blum02}) it was  
 found that the ratio $\ds |a_{\rm sc}/a_{\rm ho}|$ determines the 
 behavior. In the present case, both parameters and not 
 only their ratio are important.  
  
%----------------------------------------------------------------------
\section{Pseudopotential approximation}
%----------------------------------------------------------------------
%
\label{sec:delta}

   The bound state of two atoms in a harmonic trap when
   the atom-atom interaction $V_{\rm int}(R)$
   is approximated by a regularized contact potential 
   $\displaystyle \frac{4\pi}{2\mu} a_{\rm sc}\delta^3(\vec{R})
   \frac{\partial}{\partial R}\,R$ 
   with energy-independent scattering length $\ds a_{\rm sc}$,
   was first derived analytically by Busch {\sl at al.}~\cite{cold:busc98}.
   The bound states with integer quantum number $\ds n_t$
   are expressed as
\begin{equation}    
\displaystyle
\Psi^{n_t}_{a_{\rm sc}}(R) = \frac12\pi^{-3/2}A R e^{-\bar{R}^2/2}\Gamma
(-\nu)U(-\nu,\frac32,\bar{R}^2)\; ,
\label{WF_pseudo}
\end{equation}
   with $\bar{R}=R/a_{\rm ho}$ and the characteristic length scale 
   $a_{\rm ho}$ of the harmonic trap introduced in the end of the 
   previous section. $\ds A$ is a normalization constant 
   having the dimension of the inverse of the square root of 
   volume (see below) and $\nu$ is an effective quantum number for the 
   relative motional eigenstate,  
   $\ds \nu = \frac{E_{a_{\rm sc}}^{n_t}}{2\omega} - \frac34$.
   The energy eigenvalues are given by the roots of
   the equation
\begin{equation}  
\displaystyle
\frac{\Gamma(-x/2+3/4)}{\Gamma(-x/2+1/4)}=\frac{1}{\sqrt{2}\xi}\; ,
\label{eq:energy_pseudo}
\end{equation}
   where $x=E_{a_{\rm sc}}^{n_t}/\omega$ and $\xi=a_{\rm sc}/a_{\rm ho}$.

\begin{figure}[ht]  
 \centering
\includegraphics[width=8.5cm,height=6.5cm]{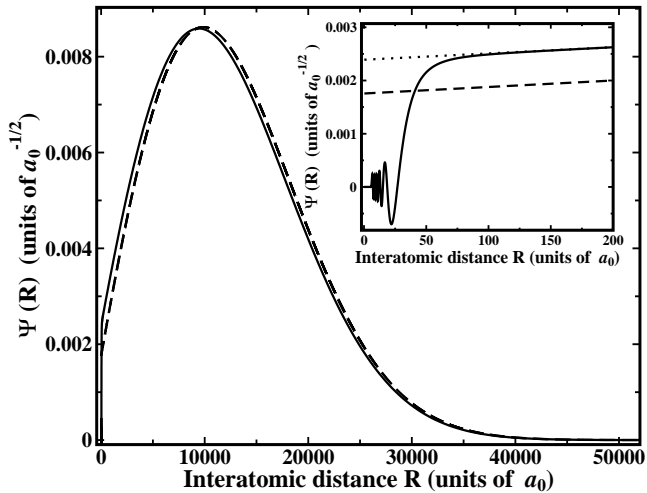}
 \caption{{\footnotesize
     Wave functions of the first trap-induced bound state 
     ($\ds \omega = 2\pi\times\mbox{10kHz}$) 
     of two $^6\mbox{Li}$ atoms interacting through the full 
     $a ^3\Sigma^+_u$ potential (solid),
     a pseudopotential with the energy-independent (trap-free) 
     scattering length $\ds a_{\rm sc} = -2030\, a_0$ (dashes), and  
     one with the energy-dependent value $\ds a_E = -2872\, a_0$ (dots). 
     The insert shows the short $R$ range on an enlarged scale. 
} }\label{fig:PPWFandRWFw10}
 \end{figure}

   The initial-state wave function $\Psi^{\rm 10'}(R;\omega)$ 
   of two $^6\mbox{Li}$ atoms interacting through the 
   $a ^3\Sigma^+_u$ potential and the pseudopotential wave 
   function $\ds \Psi_{a_{\rm sc}}^0$ with the physical (trap-free) value 
   of the scattering length $\ds a_{\rm sc}=-2030\, a_0$ are plotted 
   together in Fig.~\ref{fig:PPWFandRWFw10} for the case of a trap 
   frequency $\ds \omega = 2\pi\times\mbox{10kHz}$.
 As expected, wave function $\ds \Psi_{a_{\rm sc}}^0$
 fails completely for short internuclear separations, since 
 it does not reproduce any nodal structure at all. In 
 addition, $\ds \Psi_{a_{\rm sc}}^0$ possesses a wrong 
 behavior at $R=0$ where it is non-zero. In the long-range 
 part $\ds \Psi_{a_{\rm sc}}^0$ agrees better with the 
 correct wave function. There the main difference is an evident 
 phase shift between the two functions. This phase shift 
 is a consequence of the trap and vanishes in the absence 
 of the trap ($\omega\rightarrow 0$). The physical reason 
 for the phase shift is the non-zero ground-state energy 
 in a trap (zero-point energy and motion) due to the 
 Heisenberg uncertainty principle. As a consequence, the 
 scattering of the two atoms in a trap differs from 
 the trap-free case even at zero temperature. On the basis    
 of an analysis of the energy spectrum of two atoms in a
 harmonic trap it was found that a pseudopotential approximation 
 using an energy-dependent scattering length $\ds a_E$ leads 
 to a highly improved description of two particles confined 
 in an isotropic harmonic trap~\cite{cold:blum02,cold:bold03}.

 While the scattering length is defined in the limit 
 $E\rightarrow 0$, an energy-dependent scattering length 
 can be introduced by extending its original asymptotic  
 definition in terms of the phase shift for s-wave 
 scattering $\delta_0(E)$ to non-zero collision energies. 
 This yields $\ds a_E = -\mbox{tan} \delta_0(E)/k$ with 
 $\ds k = \sqrt{2\mu E}$. Clearly, the evaluation of 
 $\delta_0(E)$ requires to solve the complete scattering 
 problem and thus also $\ds a_E$ can only be obtained 
 from the knowledge of the solution for the 
 correct atom-atom interaction potential. 

 The values of $\ds a_E$ were obtained in the following way. 
 After a determination of the ground-state energy of two 
 $^6\mbox{Li}$ atoms from a full calculation (using the 
 realistic interaction potential), this energy is used in 
 Eq.\,(\ref{eq:energy_pseudo}) to find $\ds a_E$ 
 (that is inserted in the equation in place of 
 $\ds a_{\rm sc}$). More details about this so-called 
 self-consistency approach are given  
 in~\cite{cold:bold02}. In this way an energy-dependent 
 scattering length $\ds a_E = -2872\,a_0$ is, e.\,g., 
 found for two $^6$Li atoms in a trap with frequency 
 $\ds \omega = 2\pi\times\mbox{10kHz}$. The resulting 
 wave function is also shown in Fig.~\ref{fig:PPWFandRWFw10}, 
 together with the correct one and the one obtained for 
 $\ds a_{\rm sc}=-2030\, a_0$. Clearly, the agreement 
 with the correct wave function is very good for large 
 $R$. For $R>150\,a_0$ the wave function obtained for 
 $\ds a_E = -2872\,a_0$ is not distinguishable from 
 the correct one.   
 Only in the insert of Fig.~\ref{fig:PPWFandRWFw10} 
 that shows the wave functions at short internuclear 
 separations one sees a deviation. It is caused by 
 the absence of any nodal structure and the wrong 
 behavior at $R\rightarrow 0$ of the pseudopotential 
 wave function. In fact, at short distances the 
 introduction of an energy-dependent scattering length 
 that corrects the phase shift leads to an even 
 larger error compared to the use of $a_{\rm sc}$. 

 The validity of the pseudopotential approximation 
 using an energy-dependent scattering length has been discussed before. 
 In~\cite{cold:blum02} it was found that applicability of this approximation 
 depends on the ratios $\ds \beta_6 / a_{\rm ho}$ and 
 $\ds |a_{\rm sc} / a_{\rm ho}|$ 
 where $\ds \beta_6 = (2\mu C_6)^{1/4}$ is the characteristic 
 length scale of the interaction potential in the case of a leading 
 $C_6/R^6$ van der Waals potential. For two 
 $^6\mbox{Li}$ atoms in a trap with 
 $\ds \omega = 2\pi\times\mbox{100\,kHz}$ that interact 
 via the $\ds a^3\Sigma^+_u$ potential those ratios 
 are 0.02 and 0.59, respectively. These validity 
 criteria are, however, based solely on energy arguments. 
 In other words, if those ratios are sufficiently smaller 
 than 1, the energy obtained by means of 
 Eq.\,(\ref{eq:energy_pseudo}) with $a_{\rm sc}$ should agree well 
 with the correct one.
 In the present example of $^6\mbox{Li}$ the ratio between the 
 correct first trap-induced energy $E^{10'}$ and $E^{n_t=0}_{a_{\rm sc}}$   
 obtained with the energy-independent pseudopotential 
 is $E^{10'}/E^{0}_{a_{\rm sc}}=0.96$ for 
 $\ds \omega = 2\pi\times\mbox{10\,kHz}$ and 
 $E^{10'}/E^{0}_{a_{\rm sc}}=0.92$ for 
 $\ds \omega = 2\pi\times\mbox{100\,kHz}$. By construction, 
 the energy $E^{n_t0}_{a_{\rm E}}$ agrees of course completely 
 with $E^{10'}$. 

\begin{figure}[ht] 
 \centering
\includegraphics[width=8.5cm,height=6.5cm]{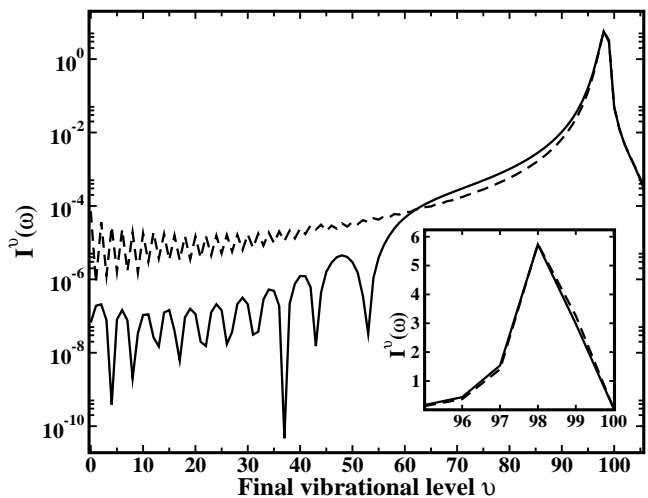}
 \caption{{\footnotesize
     Squared photoassociation transition moments $\ds I^v(\omega)$ 
     for $^6$Li in a $\omega = 2\pi\times 10\mbox{\,kHz}$ trap 
     calculated with the molecular interaction potential (solid) 
     or within the energy-independent pseudopotential approximation 
     (dashes). 
} }\label{fig:PAtransmomPPandRw10}
 \end{figure}

 In Fig.~\ref{fig:PAtransmomPPandRw10} $I^v(\omega)$ 
 obtained when using the pseudopotential approximation with 
 energy-independent scattering length is compared to the 
 spectrum obtained for the correct atom-atom interaction, both 
 for a trap frequency $\ds \omega = 2\pi\times\mbox{10\,kHz}$. 
 The two results disagree completely for $v\leq 60$. 
 For higher lying vibrational states ($v>60$) the agreement 
 is reasonable. (Note, however, the logarithmic scale.) 
 For the highest lying states ($v\ge 95$) very good agreement is found 
 even on a linear scale (see insert of Fig.~\ref{fig:PAtransmomPPandRw10}).   
 Adopting the energy-dependent scattering length yields 
 quantitative agreement already for $v\ge 75$, but again  
 a complete disagreement for $v\leq 60$. 

 The breakdown of the pseudopotential approximation (with 
 energy-independent or dependent scattering length) for describing 
 photoassociation to the low-lying vibrational states is, of course, 
 a direct consequence of the wrong short-range behavior of the 
 pseudopotential wave functions (Fig.~\ref{fig:PPWFandRWFw10}).   
 From the definition of $I^v(\omega)$ it follows that the 
 pseudopotential approximation fails, if the final-state 
 vibrational wave function has a substantial amplitude 
 in the $R$ range in which the initial-state wave function 
 is strongly influenced by the atom-atom interaction. An 
 estimate for this $R$ range is (in the present case) 
 the already discussed effective-range parameter 
 $\ds \beta_6 = (2\mu C_6)^{1/4}$. Since for large $v$ the 
 final-state wave function is dominated by its outermost lobe 
 whose position is in turn close to the classical outer 
 turning point $R_{\rm out}$, the pseudopotential 
 approximation should be valid for $\ds R_{\rm out} > \beta_6$. 
 In the case of $^6$Li one finds $\beta_6 = 62.5 \,a_0$.
 According to Fig.~\ref{fig:ClasTurnPoint} the 
 pseudopotential approximation should thus be applicable for $v>70$.
 A recently performed photoassociation experiment for 
 $ ^6 \mbox{Li}$ considered the transition to $\ds v=59$ \cite{cold:schl03}. 
 For this specific example the pseudopotential approximation 
 would predict a two times smaller rate than the full calculation 
 for $\ds \omega = 2\pi\times\mbox{10\,kHz}$. %

 The validity of the pseudopotential approximation for 
 predicting the photoassociation rates to the high lying 
 states can also be used to investigate whether the 
 simulation of different scattering lengths by mass scaling 
 is senseful. This could be questionable, since a change of the 
 mass does not only modify the scattering length (by moving the 
 position of the least bound state), but also the kinetic 
 energy term. Thus it may be argued that the discussed 
 influence of the scattering length on the photoassociation 
 process could partly also be a consequence of the 
 modification of the kinetic energy, at least in the case of a 
 substantial mass variation as it was required for preserving the 
 number of bound states. Within the pseudopotential 
 approximation the scattering length is, however, a parameter 
 independent of the mass. Therefore, in contrast to the case of 
 the full calculation it is possible within the pseudopotential 
 approximation to investigate the isolated influence of a variation 
 of the scattering length (keeping the mass fixed). A corresponding 
 analysis confirms that mass scaling can in fact be used to 
 simulate a modified atom-atom interaction. 

\begin{figure}[ht] 
 \centering
\includegraphics[width=8.5cm,height=13cm]{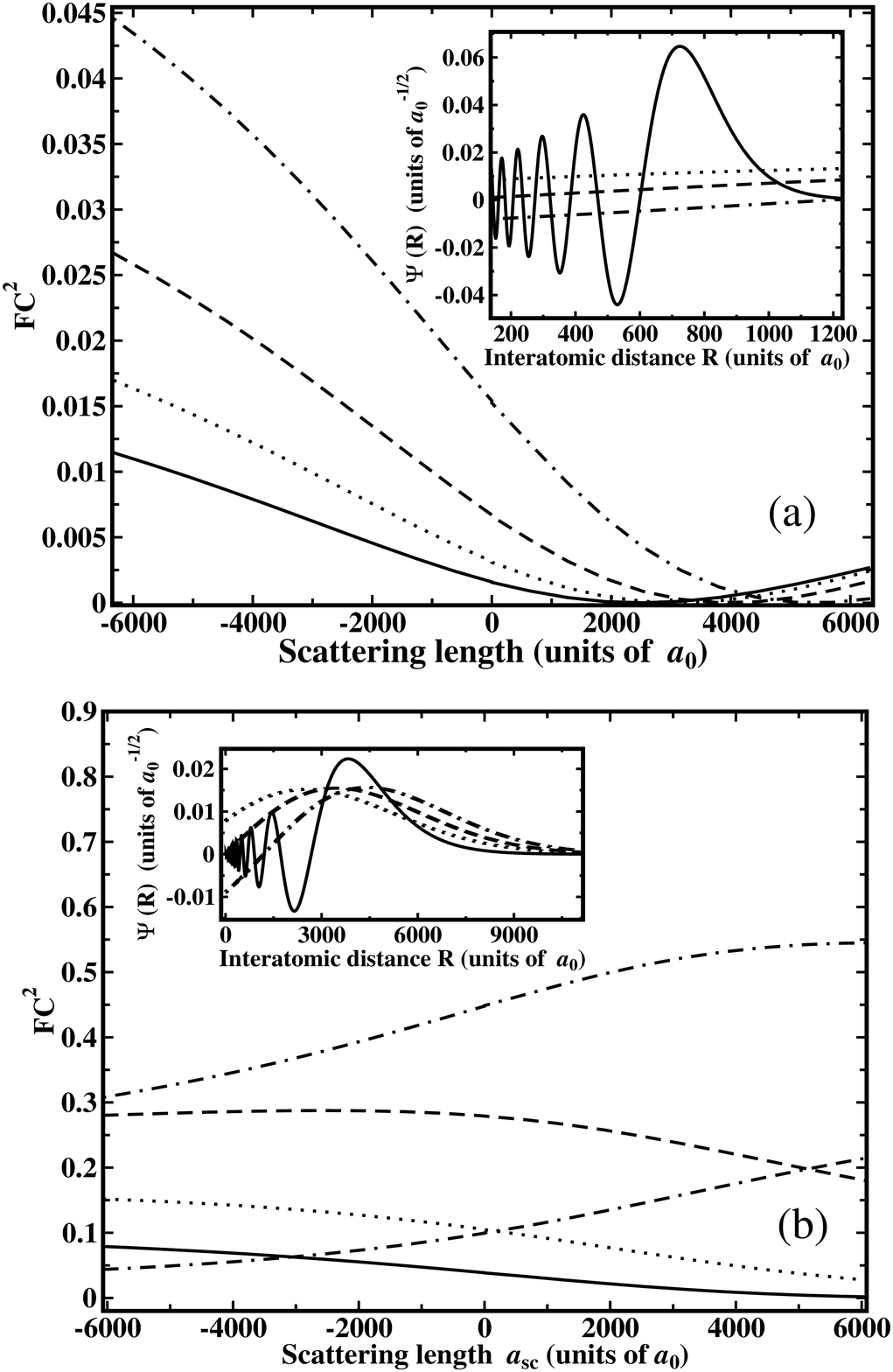}
 \caption{{\footnotesize
     (a) Squared Franck-Condon factors between the final vibrational states 
     $v=90$ (solid), 91 (dots), 92 (dashes), or 93 (chain) 
     of the $1 ^3\Sigma^+_g$ state and the initial-state pseudopotential 
     wave functions as a function of the scattering length $a_{\rm sc}$.  
     The trap frequency is $\ds \omega = 2\pi\times\mbox{100\,kHz}$. 
     The insert shows the $v=92$ final-state wave function 
     together with the pseudopotential wavefunctions 
     for $a_{sc}=-2000\, a_0$ (dots), 
     $a_{sc}=0$ (dashes), $a_{sc}=+2000\, a_0$ (chain).   
     (b) As (a), but for $v=94$ (solid), 95 (dots), 96 (dashes),
     97 (chain), v=98 (dot-dash-dash). The insert shows the 
     $v=97$ final-state wavefunction and the pseudopotential 
     wave functions for $a_{sc}=-2000\, a_0$ (dots), 
     $a_{sc}=0$ (dashes), $a_{sc}=+2000\, a_0$ (chain).
} }\label{fig:FCascvar}
 \end{figure}

 The pseudopotential approximation was used already in~\cite{cold:deb03} 
 for an analysis of the change of the photoassociation rate due to 
 a scattering-length modification. The investigation concentrated, however, 
 on very high lying vibrational states close to or even above the 
 trap-free dissociation limit. Since for transitions to those states 
 the $R$ dependence of the electronic transition dipole moment can 
 safely be ignored, it is sufficient to concentrate on the Franck-Condon 
 (FC) factors. In Fig.~\ref{fig:FCascvar} the squares of these factors 
 are shown as a function of the scattering length for $90\le v\le 98$ and 
 trap frequency $\ds \omega = 2\pi\times\mbox{100\,kHz}$.
 As in~\cite{cold:deb03} the pseudopotential approximation is used for 
 the initial state, but here the final-state wave function is obtained 
 by a full numerical calculation while an approach based on quantum defect 
 theory (QDT) was used in~\cite{cold:deb03}. Furthermore, Na$_2$ was 
 considered in~\cite{cold:deb03} while it is Li$_2$ in the present study.  

 For the states $90\le v \le 93$ shown in Fig.~\ref{fig:FCascvar}(a) 
 the dependence on $a_{\rm sc}$ in a 100\,kHz trap is very similar to the 
 one found in~\cite{cold:deb03}. The rather regular variation with 
 $a_{\rm sc}$ is due to the fact that the final-state wave function 
 probes the flat part of the initial-state wave function, as can be 
 seen in the insert of Fig.~\ref{fig:FCascvar}(a) where the wave function 
 for $v=92$ is shown together with the initial-state wave function for 
 three different values of $a_{\rm sc}$. The initial-state wave function 
 varies almost linearly with $a_{\rm sc}$ in the Franck-Condon window 
 of the $v=92$ final state. According to the discussion in 
 Sec.\,\ref{sec:cutoffregim}, for a 100\,kHz trap the states $v\ge 90$ 
 belong to the cut-off regime, but for $v\le 93$ the enhancement factor 
 $f^v$ is still close to its value $f_c$ in the constant regime 
 (see Figs.~\ref{fig:PAtransmomAttract} and~\ref{fig:PAtransmomRepuls}).
 The minima of the FC$^2$ factors for $a_{\rm sc}\gg 0$ are a consequence 
 of the dip discussed in Sec.\,\ref{sec:positivesclen}. Since the 
 nodal position $R_{\rm x}$ moves to larger $R$ if $a_{\rm sc}$ 
 increases, the minimum in the FC$^2$ factors moves to a larger value 
 of $a_{\rm sc}$ if $v$ increases. While the pseudopotential approximation 
 is capable to predict the existence of the dip occuring for 
 $a_{\rm sc}\gg 0$, its position is not necessarily correctly reproduced 
 in a trap. This is due to the fact that the pseudopotential overestimates 
 the trap-induced shift of the position of the outermost node. 
 For example, if the mass of Li is varied such that 
 $a_{\rm sc} = + 850\,a_0$ is obtained, a 100\,kHz trap shifts $R_{\rm x}$ 
 to $\approx +810\,a_0$ (Fig.\,\ref{fig:dip}) and the dip occurs at $v=92$    
 (Fig.\,\ref{fig:PAtransmomRepuls}). Using the pseudopotential approximation 
 (with $a_{\rm sc}=+850\,a_0$) yields on the other hand 
 $R_{\rm x}\approx +580\,a_0$ and the dip occurs for $v=90$. This error in 
 the prediction of $R_{\rm x}$ increases with $a_{\rm sc}$. 
  
 The final states $94\le v \le 98$ 
 whose FC$^2$ factors are shown in Fig.~\ref{fig:FCascvar}(b) probe on 
 the other hand the non-linear part of the initial-state wave function 
 (close to the trap boundary). Consequently, the dependence on $a_{\rm sc}$ 
 differs from the one found in~\cite{cold:deb03}. While for 
 $90\le v \le 92$ the FC$^2$ factors are first decreasing and then increasing, 
 if $a_{\rm sc}$ varies from $-6000\,a_0$ to $+6000\,a_0$, the ones of 
 $93\le v \le 96$ are purely decreasing. For $v=97$ and 98 the FC$^2$ 
 factors are on the other hand increasing with $a_{\rm sc}$.

 In view of the fact that the scattering length of a 
 given atom pair may be known (for example from some 
 measurement), but the corresponding atom-atom 
 interaction potential is unknown, it is of course 
 interesting to investigate whether the pseudopotential 
 approximation allows to predict the enhancement 
 factor also in the constant regime, i.\,e.\ whether 
 it correctly reproduces $f_c(\omega)$. This 
 would allow for a simple estimate of the effect of a 
 tight trap on the photoassociation rate in the constant 
 regime that covers most of the spectrum. In order to 
 determine $f_c(\omega)$ it is sufficient to analyze the 
 ratio of the initial-state wave function $\Psi_{a_{\rm sc}}^0$ 
 for the trap frequencies $\omega$ and $\omega_{\rm ref}$. 
 This comparison may be done at any arbitrary internuclear 
 separation $R_{\rm lin}$ provided it belongs to the linear 
 regime. The result is %
\begin{equation} 
\ds
  \ds f_c^{\rm pseudo}({\omega}) 
          = \left[\frac{\Psi^{0}_{a_{\rm sc}}(R_{\rm lin};\omega)}
                       {\Psi^{0}_{a_{\rm sc}}
                           (R_{\rm lin};\omega_{\rm ref})}\right]^2\; .
\label{rule1a}
\end{equation}

 A very special and simple choice that guarantees that 
 $R_{\rm lin}$ belongs to the linear regime is $R_{\rm lin}=0$. 
 With this value of $R_{\rm lin}$ one finds from the analysis 
 of $\Psi^{0}_{a_{\rm sc}}$ % 
\begin{equation}  
\ds
  \ds f_c^{\rm pseudo}({\omega}) = 
          \left[\frac{A({\omega})}{A({\omega_{\rm ref}})}\right]^2
          \: \frac{\omega_{\rm ref}}{\omega}\; ,
\label{eq:rule1b}
\end{equation}
 where $\ds A({\omega})$ is the normalization factor fulfilling
 $\ds |A({\omega})|^2 \,
       =\,\sqrt{2\omega}\,\pi\,\xi^2 \,\frac{\partial E}{\partial\xi}$%
~\cite{cold:busc98}. Depending on the level of approximation 
 one may use either $a_{\rm sc}$ or $a_E$ in the evaluation of 
 $A$. 
 An even simpler estimate of the influence of a tight harmonic 
 trap on the photoassociation rate is obtained, if the atom-atom 
 interaction potential is completely ignored in the initial state. 
 The harmonic-oscillator eigenfunctions at $R=R_{\rm lin}=0$ 
 yield %
\begin{equation}  
   f_c^{\rm ho}({\omega}) = \left(\,
                                \frac{\omega}{\omega_{\rm ref}}   
                            \right)^{3/2} \; .
\label{ruleharm}
\end{equation}

\begin{figure}[ht]     
 \centering
\includegraphics[width=8.5cm,height=6.5cm]{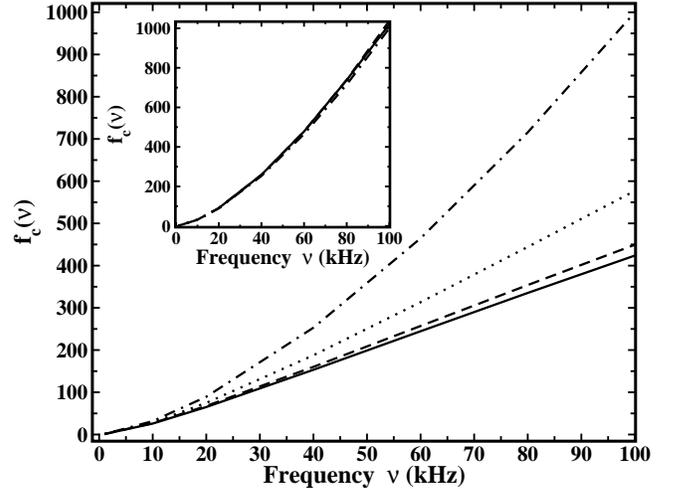}
 \caption{{\footnotesize
     Dependence of the enhancement factor $\ds f_{c}$ on the trap 
     frequency $\nu=\omega/(2\pi)$ for $ ^6 \mbox{Li}_2$ using 
     the molecular (solid), the energy-independent (dots) and 
     -dependent (dashes) pseudopotential, or the harmonic-oscillator (chain)  
     wave functions. The insert shows the same curves but for
     $^{39}$K$_2$.
} }\label{fig:PAratiowvar}
 \end{figure}

 In Fig.~\ref{fig:PAratiowvar} the enhancement factors $f_c(\omega)$ 
 calculated at the different levels of approximation are shown as 
 a function of the trap frequency $\omega$.
 The results obtained for 
 $^6$Li and $^{39}$K are compared to each other. Remind, in the latter 
 case the scattering length $a_{\rm sc}= +90\,a_0$ has a much 
 smaller absolute value than for $^6$Li ($a_{\rm sc}=-2030\,a_0$). 
 Consequently, one expects the atom-atom interaction to be less 
 important. This is confirmed by (the insert of) 
 Fig.~\ref{fig:PAratiowvar}. The results obtained for $f_c(\omega)$ 
 with the aid of the different approximations discussed above 
 are in very good agreement with the correct result in the case 
 of $^{39}$K. Even the simple harmonic-oscillator model predicts 
 the enhancement factor in the constant regime very accurately. 

 It should be emphasized that the correct prediction of the enhancement 
 factor by the simplified approximation works, although the prediction of 
 the rates is completely wrong (Fig.\,\ref{fig:PAtransmomPPandRw10})  
 in this constant regime (small $v$). In the case of a large absolute 
 value of the scattering length (like for $^6$Li), i.\,e.\ for a strong 
 atom-atom interaction, the frequency dependence of $f_c(\omega)$ 
 predicted by the simplified models is on the other hand not very 
 accurate. In fact, the simple harmonic-oscillator model clearly 
 overestimates the enhancement factor for large $\omega$. 
 The pseudopotential approximation yields much better results, 
 especially if the energy-dependent scattering length $a_E$ is 
 used. (As already mentioned, $a_E$ is, however, only available from 
 the knowledge of the exact atom-atom interaction.)

\begin{figure}     
 \centering
\includegraphics[width=8.5cm,height=6.5cm]{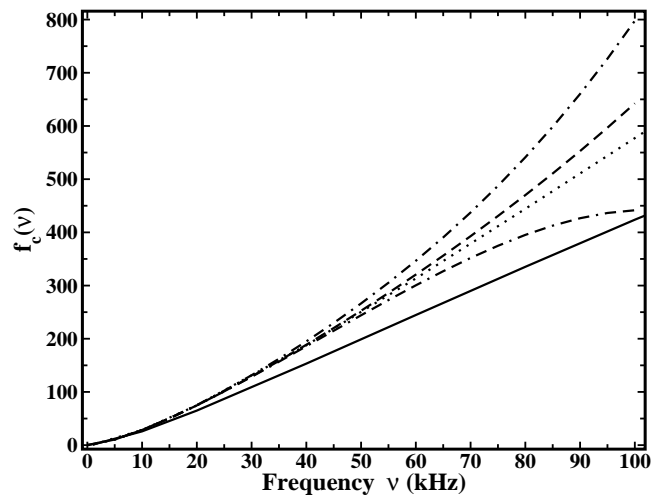}
 \caption{{\footnotesize
     Investigation of the convergence of the series 
     expansion\,(\ref{eq:rule1c}) for the enhancement factor  
     $\ds f_{\rm c}^{\rm pseudo}(\omega)$ in the constant regime. 
     Shown are the results of the full (infinite order, dotted)  
     and the 4th (dashes), 5th (chain), and 6th (dot-dash-dash) 
     order expansion. For comparison, also the enhancement 
     factor obtained with the molecular potential is shown (solid).  
} }\label{fig:simpleestim}
 \end{figure}

 In view of the usefulness of Eq.~(\ref{eq:rule1b}) for obtaining an 
 estimate of the enhancement factor $f_c(\omega)$ but the rather 
 complicated procedure to calculate $\ds \frac{\partial E}{\partial\xi}$ 
 required for obtaining $\ds A({\omega})$, it is interesting to 
 test whether $\ds A({\omega})$ can alternatively be evaluated 
 from an expansion of the energy $\ds E$ at $\ds \xi=0$.
 Using the relation  
 $\ds \frac{\partial x}{\partial \xi} = 
                 \left(\frac{\partial \xi}{\partial x}\right)^{-1}$
 it is straightforward to determine with the aid 
 of~(\ref{eq:energy_pseudo}) an expansion for the scaled energy
 \begin{equation} 
   \ds
      x(\xi) = \frac32+\sum\limits_{n=0}^{\infty}
                   \frac{1}{(n+1)!} \left.
                   \frac{\partial^{(n)} F(x)}{\partial x^{(n)}}
                   \right|_{x=3/2} \xi^{n+1} \,
   \label{eq:energy_expansion}
 \end{equation}
 with 
 \begin{equation} 
   \ds F(x) = -\frac{2\sqrt{2}\,\Gamma\Bigl[\frac34-\frac{x}2\Bigr]}
                    {\Gamma\Bigl[\frac14-\frac{x}2\Bigr]
                     \psi\Bigl[\frac14-\frac{x}2\Bigr]-
                     \psi\Bigl[\frac34-\frac{x}2\Bigr]}\, 
 \end{equation}
 and the digamma function $\psi$. The zero- and first-order terms
 of the expansion (\ref{eq:energy_expansion}) are given
 in~\cite{cold:busc98}. Using Eqs.~(\ref{eq:energy_expansion})
 and~(\ref{eq:rule1b}) %
  \begin{equation} 
   \ds f_c^{\rm pseudo}({\omega}) =\frac{\sum\limits_{n=0}^{\infty}
                   \frac{1}{n!} \left.
                   \frac{\partial^{(n)} F(x)}{\partial x^{(n)}}
                   \right|_{x=3/2} \xi^{n}}{\sum\limits_{n=0}^{\infty}
                   \frac{1}{n!} \left.
                   \frac{\partial^{(n)} F(x)}{\partial x^{(n)}}
                   \right|_{x=3/2} \xi^{n}_{\rm ref}}
                   \Bigl(\frac{\omega}{\omega_{\rm ref}}\Bigr)^{\frac32} 
        \label{eq:rule1c}
 \end{equation}
 is obtained with $\ds \xi_{\rm ref}=\frac{a_{\rm sc}}{a_{\rm ho, ref}}$.
 In Fig.~\ref{fig:simpleestim} the 4th-, 5th-, and 6th-order 
 expansions are compared to the results obtained with the 
 non-approximated term (all for 
 $a_{\rm sc}=-2030 \, a_0$) and with the correct atom-atom interaction 
 result.
 Note, Eq.~(\ref{eq:rule1c}) can also be used for the evaluation of 
 $g_c(\omega,a_{\rm sc})$, if in the denominator $\xi_{\rm ref}$ 
 is replaced by 
 $\ds \tilde{\xi}_{\rm ref}=\frac{a_{\rm sc,ref}}{a_{\rm ho, ref}}$.

%-------------------------------
\section{Discussion and Outlook}
%-------------------------------
%
\label{sec:discussion}

 In this work the influence of a tight isotropic harmonic 
 trap on the photoassociation process has been investigated 
 for alkali atoms. It is found that for most of the states 
 (the ones in the constant regime) there is an identical  
 enhancement as the trap frequency increases. This enhancement 
 can reach 3 orders of magnitude for trap frequencies of about 
 100\,kHz as they are reported in literature. While the 
 enhancement itself agrees at least qualitatively with the 
 concept of confinement of the initial-state wave function, 
 also trap-induced suppressed photoassociation is possible. 
 In fact, as a simple sum rule confirms, any enhancement 
 must be accompanied by suppression. The physical origin 
 of this suppression is the trap-induced confinement of the initial-state 
 wave function of relative motion within a radius that is smaller 
 than the mean internuclear separation of the least bound vibrational 
 states in the electronic target state. Since in the present calculation 
 both initial and final state are exposed to the same harmonic 
 trap, this result may appear surprising. While the explanation is 
 based on the different long-range behaviors of the two involved 
 electronic states, the effect itself may be very interesting in 
 terms of, e.\,g., quantum information. 

 Consider for example an optical lattice 
 as trapping potential. The initial (unbound) atom pair is 
 (for sufficient trap depths) located within a single 
 lattice site (Mott insulator state). In the photoassociated 
 state it could, however, reach into and thus communicate with 
 the neighbor site, if the lattice parameters are appropriately 
 chosen. Such a scenario could be used for a controlled logical 
 operation (two-qubit gate) like the CNOT. Since the latter forms  
 together with single-qubit gates a universal gate, this could 
 provide a starting point for a quantum computer. Alternatively, 
 it may be interesting to use the fact that if a single spot with  
 the dimension of the trap length $a_{\rm ho}$ or a specific 
 site in an optical lattice can be addressed, then the 
 atoms would only respond, if they are in their (unbound) initial 
 state. If they are in the photoassociated excited state, they 
 would on the other hand be located outside the trap and thus 
 would not respond. For this it is already sufficient, if they 
 are (predominantly) located in the classically forbidden regime.    
 Also, modifying the trap frequency it is possible to block the 
 photoassociation process on demand. The trap frequency is then  
 varied in such a fashion that a specific final state resonantly 
 addressed with a laser with sufficiently small bandwidth belongs 
 either to the constant or to the cut-off regime. 

 A further important finding of this work is that the influence of a 
 tight trap on the photoassociation spectra (as a function of the final 
 vibrational state) for different alkali atoms is structurally 
 very similar, independent whether photoassociation starts from the 
 singlet or triplet ground state. Also the type of interaction (strong or 
 weak as well as repulsive or attractive) does not lead to a substantial 
 modification of the trap influence. The only exception is a strong 
 repulsive interaction that leads to a pronounced window in the 
 photoassociation spectrum. The reason is the position of the 
 last node in the initial-state wave function that in this case 
 is located at a relatively large value of $R$ and leads to a 
 cancellation effect in the overlap with the final state. The nodal 
 position depends strongly on $a_{\rm sc}$, but only for very tight 
 traps also on $\omega$. As has been 
 discussed previously~\cite{cold:cote98b,cold:abra96}, the position of 
 the window may be used for a scattering-length determination. 
 This will also approximately work for not too tight traps, but the trap 
 influence has to be considered for very tight ones. Alternatively, the 
 window provides a control facility, since the transition to a single 
 state can be selectively suppressed. In very tight traps this effect 
 is not only more pronounced, but in addition the transitions to the 
 neighbor states are further enhanced. This could open up a new road 
 to control in the context of the presently on-going discussion of 
 using femtosecond lasers for creating non-stationary wave packets in 
 the electronic excited state~\cite{cold:salz06,cold:brow06,cold:koch06}. 
 One of the problems encountered in this approach is the difficulty to 
 shape the wave packet, since the high-lying vibrational states that have a 
 reasonable transition rate are energetically very closely spaced 
 and thus the shape of the wavepacket is determined by the 
 Franck-Condon factors that cannot easily be manipulated but strongly 
 increase as a function of $v$. 

 In view of the question how to enhance photoassociation or related 
 association schemes (like Raman-based ones) the investigation of the 
 enhancement factors $g^{v}(\omega,a_{\rm sc})$, especially its value 
 in the constant regime ($g_c(\omega,a_{\rm sc})$) is most important. 
 It shows that not only increasing the tightness of the trap (enlarging 
 $\omega$) leads to an enhancement of the photoassociation rate, but 
 a similar effect can be achieved by increasing the interaction 
 strength $|a_{\rm sc}|$. Most interestingly, these two enhancement 
 factors work practically independent of each other, i.\,e.\ it is 
 possible to use both effects in a constructive fashion and to obtain 
 a multiplicative overall enhancement factor. For a 100\,kHz trap and 
 a scattering length $|a_{\rm sc}|$ of the order of 2000 an enhancement 
 factor (uniform for all states in the constant regime) of 5 to 6 
 orders of magnitude is found compared to the case of a shallow 
 1\,kHz trap and $|a_{\rm sc}|= 0$.  

 A comparison of the results obtained for the correct atom-atom 
 interaction potential with the ones obtained using the approximate 
 pseudopotential approximation or ignoring the interaction at all 
 shows that these approximations yield only for the transitions to 
 very high lying vibrational states a good estimate of the photoassociation 
 rate. Nevertheless, despite the complete failure of predicting the 
 rates to low lying states, these models allow to determine the 
 enhancement factor in the constant regime. For weakly interacting 
 atoms (small $|a_{\rm sc}|$) already the pure harmonic-oscillator 
 model (ignoring the atomic interaction) leads to a reasonable 
 prediction of the trap-induced enhancement factor $f_c(\omega)$. 

 It is important to stress that the results in this work were obtained 
 for isotropic harmonic traps with the same trapping potential seen by 
 both atoms. In this case center-of-mass and relative motion can be 
 separated and in both coordinates an isotropic harmonic trap potential 
 (with different trap lengths due to the different total and reduced 
 masses) is encountered. 
 As is discussed, e.\,g., in \cite{cold:bold03,cold:idzi05} where a 
 numerical and an analytical solution are respectively derived for the 
 case of atoms interacting via a pseudopotential, a similar 
 separation of center-of-mass and relative motion is possible for 
 axially symmetric (cigar or pancake shaped) harmonic traps.       

 In reality, the traps for alkali atoms are of course not strictly 
 harmonic. Since the present work focuses, however, for the initial 
 atom pair on the lowest trap induced state the harmonic approximation 
 should in most cases be well justified. Independently on the exact 
 way the trap is formed (for example by a far off-resonant focused 
 Gaussian laser beam or by an optical lattice), the lowest trap-induced 
 state agrees usually well with the one obtained in the harmonic 
 approximation, 
 if the zero-point energy is sufficiently small. This requirement 
 sets of course an upper scale to the applicability of the harmonic 
 approximation with respect to the trap frequency. If $\omega$ is too 
 large, the atom pair sees the anharmonic part of the trap. (Clearly, 
 the trap potential must also be sufficiently deep to support trap-induced 
 bound states, i.\,e.\ to allow for Mott insulator states in the case of 
 an optical lattice). 

 An additional problem arises from the 
 anharmonicity of a real trap: the anharmonic terms lead to a 
 non-separability of the relative and the center-of-mass motion. 
 In fact, a recent work discusses the possibility of using this coupling 
 of the two motions for the creation of molecules \cite{cold:bold05}.  
 Again, a tighter trap is expected to lead to a stronger coupling 
 and thus finally to a breakdown of the applicability of the harmonic 
 model. 

 For the final state of the considered photoassociation process there 
 exists on the first glance an even more severe complication. Usually, 
 the two atoms will not feel the same trapping potential, since they 
 populate different electronic states. In the case of traps whose 
 action is related to the induced dipole moment (which is the case 
 for optical potentials generated with the aid of lasers that are 
 detuned from an atomic transition), the two atoms (in the 
 case of Li the ones in the 2\,$^2$S and the 2\,$^2$P state) will in 
 fact see potentials with opposite sign. If the laser traps the ground-state 
 atoms, it repels the excited ones. In the alternative case of an 
 extremely far-off resonant trap the trapping potential is proportional 
 to the dynamic polarizability of the atoms. In the long-wavelength limit 
 (as is realized, e.\,g., in focused CO$_2$ lasers~\cite{cold:take95}) 
 the dynamic polarizability approaches the static one, 
 $\lim_{\lambda\rightarrow \infty} \alpha(\lambda) = \alpha_{\rm st}$.      
 The static polarizabilities do not necessarily have opposite signs 
 for the ground and the excited electronic state of an alkali atom, 
 but in many cases different values. Then the trapping potentials  
 for the initial and final states of the photoassociation process 
 are different. The Li system appears to be a counter example, since 
 for $ ^6 \mbox{Li}_2$ the average polarizability of the 
 $a ^3\Sigma^+_u$ (2s+2s) state is predicted to be equal to 
 $\overline{\alpha}=\alpha_{zz}=\alpha_{xx}=2\alpha_0 (2s) = 
 2\times165= 330 \,a_0$ For the $1 ^3\Sigma^+_g$ (2s+2$p_z$) state 
 one has  
 $\alpha_{zz}=\alpha_0(2s)+\alpha_{zz}(2p_z)=285\, a_0$ and 
 $\alpha_{xx}=\alpha_0(2s)+\alpha_{xx}(2p_z)=292 \, a_0$
 yielding an average polarizability $\overline{\alpha} 
 \approx 290 \,a_0$~\cite{cold:mera01}. Thus the trapping 
 potentials are expected to be very similar. This is 
 not the case for, e.\,g., $^{87}$Rb$_2$ where the average 
 polarization for the $a ^3\Sigma^+_u$ state is $670 \, a_0$ and
 for $1 ^3\Sigma^+_g$ it is $1698 \,a_0$~\cite{cold:magn02}.
 
 It was checked that the use of very different values of 
 $\omega$ for determining the initial and final state wave functions 
 does not influence the basic findings of the present work. The reason 
 is simple. Besides the very least bound states (and of course the 
 trap-induced ones) the final states are effectively protected 
 by the long-range interatomic potential from seeing the trap.     
 However, if the two atoms are exposed to different trap 
 potentials, a separation of center-of-mass and relative motion 
 is again not possible, even in the fully harmonic case (a fact that 
 was, e.\,g., overlooked in~\cite{cold:koch05}). One 
 would again expect that this coupling increases with the 
 difference in the trap potentials of the two involved states. 
 A detailed study of the consequences of the coupling of 
 center-of-mass and relative motion due to various reasons 
 like an anharmonicity of the trap or different trapping potentials 
 seen by the involved atoms is presently underway. This involves 
 also the case of the formation of heteronuclear alkali dimers 
 where besides the occurrence of this coupling for the initial 
 state also a different long-range behavior of the interatomic 
 potential has to be considered. 

 Different interaction strengths occur naturally for different 
 alkali atoms as is well known and also evident from the explicit 
 examples of $^6$Li, $^7$Li, and $^{39}$K that were discussed in this 
 work. According to the findings of this work the choice of a proper 
 atom pair (with large $|a_{\rm sc}|$) enhances the achievable 
 photoassociation yield quite dramatically. Clearly, for practical 
 reasons it is usually not easy to change in an existing experiment 
 the atomic species, since the trap and cooling lasers are adapted 
 to a specific one. In addition, the naturally existing alkali species 
 provide only a fixed and limited number of interaction strengths. 

 The tunability of the interaction strength on the basis of Feshbach 
 resonances, especially magnetic ones, marked a very important 
 corner-stone in the research area of ultracold atomic gases. The 
 findings of the present work strongly suggest that this tunability 
 could be used to improve the efficiency of photoassociation 
 (and related) schemes. However, it has to be emphasized that it is 
 not at all self-evident that the independence of the scattering-length 
 variation and the one of the trap frequency as it occurs in the 
 model used in this work is applicable to (magnetic) Feshbach 
 resonances. Furthermore, the present work considered only the 
 single-channel case while the proper description of a magnetic Feshbach 
 resonance requires a multi-channel treatment. Noteworthy, a strong 
 enhancement of the photoassociation rate by at least 2 orders of 
 magnitude while scanning over a magnetic Feshbach resonance was 
 predicted on the basis of a multichannel calculation for a specific 
 $^{85}$Rb resonance already in~\cite{cold:abee98}. An experimental 
 confirmation followed very shortly thereafter~\cite{cold:cour98}. 
 The explanation for the enhancement given in~\cite{cold:abee98} is, 
 however, based on an increased admixture of bound-state contribution 
 to the initial continuum state in the vicinity of the resonance. This 
 is evidently different from the reason for the enhancement due to 
 large values of $|a_{\rm sc}|$ discussed in the present work.

 An extension within the multichannel formalism that 
 allows for a full treatment of magnetic Feshbach resonances is presently 
 underway.

%-------------------------
\section*{Acknowledgments}
%-------------------------
%

 The authors acknowledge financial support by the {\it Deutsche
 Forschungsgemeinschaft} within the SPP\,1116 (DFG-Sa\,936/1) and SFB\,450. 
 AS is grateful to the {\it Stifterverband f\"ur die Deutsche Wissenschaft} 
 (Programme {\it Forschungsdozenturen}) and the {\it Fonds der Chemischen 
 Industrie} for financial support. This work is also supported by the 
 European COST Programme D26/0002/02.

%\bibliography{cold,aies,bsp}

\end{document}